\RequirePackage[left,columnwise]{lineno}
\documentclass[aps,prc,twocolumn,floatfix,superscriptaddress,nofootinbib]{revtex4-1}
\usepackage{lipsum}
\usepackage{graphicx}
\usepackage{subfigure}
\usepackage{amsmath,amssymb,amsfonts,eufrak,mathrsfs}
\usepackage{bm}% bold math
\usepackage{color}
\usepackage[bookmarks,
                bookmarksopen = true,
                bookmarksnumbered = true,
                linktocpage,
                colorlinks = true,
                linkcolor = blue,
                urlcolor  = blue,
                citecolor = blue,
                anchorcolor = green,
                hyperindex = true,
                hyperfigures]
                {hyperref}
\usepackage{lineno}

\newcommand{\trento}{T\raisebox{-.5ex}{R}ENTo }
\newcommand{\be}{\begin{equation}}
\newcommand{\ee}{\end{equation}}

\newcommand{\bea}{\begin{eqnarray}}
\newcommand{\eea}{\end{eqnarray}}

\begin{document}

\title{Towards the determination of heavy-quark transport coefficients in quark-gluon plasma}

\author{Shanshan Cao}
\email[Correspondence: ]{xnwang@lbl.gov, sshan.cao@gmail.com}
\affiliation{Department of Physics and Astronomy, Wayne State University, Detroit, Michigan 48201, USA}

\author{Gabriele Coci}
\affiliation{Department of Physics and Astronomy, University of Catania, Via Santa Sofia 64, 1-95125 Catania, Italy}
\affiliation{Laboratori Nazionali del Sud, INFN-LNS, Via Santa Sofia 62, I-95123 Catania, Italy}

\author{Santosh Kumar Das}
\affiliation{School of Physical Science, Indian Institute of Technology Goa, Ponda, Goa, India}
\affiliation{Department of Physics and Astronomy, University of Catania, Via S. Sofia 64, 1-95125 Catania, Italy}

\author{Weiyao Ke}
\affiliation{Department of Physics, Duke University, Durham, North Carolina 27708, USA}

\author{Shuai Y.F. Liu}
\affiliation{Cyclotron Institute and Department of Physics and Astronomy, Texas A\&M University, College Station, Texas 77843, USA}

%\author{Tan Luo}
%\affiliation{Key Laboratory of Quark and Lepton Physics (MOE) and Institute of Particle Physics, Central China Normal University, Wuhan 430079, China}

\author{Salvatore Plumari}
\affiliation{Department of Physics and Astronomy, University of Catania, Via S. Sofia 64, 1-95125 Catania, Italy}

\author{Taesoo Song}
\affiliation{Institut f\"{u}r Theoretische Physik, Universit\"{a}t Gie\ss en, Germany}

\author{Yingru Xu}
\affiliation{Department of Physics, Duke University, Durham, North Carolina 27708, USA}

\author{J\"{o}rg Aichelin}
\affiliation{SUBATECH, IMT Atlantique, Universit\'{e} de Nantes, CNRS-IN2P3, Nantes, France}

\author{Steffen Bass}
\affiliation{Department of Physics, Duke University, Durham, North Carolina 27708, USA}

\author{Elena Bratkovskaya}
\affiliation{Institute for Theoretical Physics, Johann Wolfgang Goethe Universit\"{a}t, Frankfurt am Main, Germany}
\affiliation{GSI Helmholtzzentrum f\"{u}r Schwerionenforschung GmbH, Darmstadt, Germany}

\author{Xin Dong}
\affiliation{Nuclear Science Division, Lawrence Berkeley National Laboratory, Berkeley, California 94740, USA}

\author{Pol Bernard Gossiaux}
\affiliation{SUBATECH, IMT Atlantique, Universit\'{e} de Nantes, CNRS-IN2P3, Nantes, France}

\author{Vincenzo Greco}
\affiliation{Department of Physics and Astronomy, University of Catania, Via S. Sofia 64, 1-95125 Catania, Italy}
\affiliation{Laboratori Nazionali del Sud, INFN-LNS, Via S. Sofia 62, I-95123 Catania, Italy}

\author{Min He}
\affiliation{Department of Applied Physics, Nanjing University of Science and Technology, Nanjing 210094, China}

\author{Marlene Nahrgang}
\affiliation{SUBATECH, IMT Atlantique, Universit\'{e} de Nantes, CNRS-IN2P3, Nantes, France}

%\author{Guang-You Qin}
%\affiliation{Key Laboratory of Quark and Lepton Physics (MOE) and Institute of Particle Physics, Central China Normal University, Wuhan 430079, China}

\author{Ralf Rapp}
\affiliation{Cyclotron Institute and Department of Physics and Astronomy, Texas A\&M University, College Station, Texas 77843, USA}

\author{Francesco Scardina}
\affiliation{Department of Physics and Astronomy, University of Catania, Via S. Sofia 64, 1-95125 Catania, Italy}
\affiliation{Laboratori Nazionali del Sud, INFN-LNS, Via S. Sofia 62, I-95123 Catania, Italy}

\author{Xin-Nian Wang}
\email[Correspondence: ]{xnwang@lbl.gov, sshan.cao@gmail.com}
\affiliation{Key Laboratory of Quark and Lepton Physics (MOE) and Institute of Particle Physics, Central China Normal University, Wuhan 430079, China}
\affiliation{Nuclear Science Division, Lawrence Berkeley National Laboratory, Berkeley, California 94740, USA}

\date{\today}

%%%%%%%%%%%%%%%%%%%%%%%%%%%%%%%%%%%%%%%%%%%%%%%%%%%%%%%%%%%%%%%%%%%%%

\begin{abstract}

Several transport models have been employed in recent years to analyze heavy-flavor meson spectra in high-energy heavy-ion collisions. Heavy-quark transport coefficients extracted from these models with their default parameters vary, however, by up to a factor of 5 at high momenta. To investigate the origin of this large theoretical uncertainty, a systematic comparison of heavy-quark transport coefficients is carried out between various transport models. Within a common scheme devised for the nuclear modification factor of charm quarks in a brick medium of a quark-gluon plasma, the systematic uncertainty of the extracted drag coefficient among these models is shown to be reduced to a factor of 2, which can be viewed as the smallest intrinsic systematical error band achievable at present time. This indicates the importance of a realistic hydrodynamic evolution constrained by bulk hadron spectra and of heavy-quark hadronization for understanding the final heavy-flavor hadron spectra and extracting heavy-quark drag coefficient.  The transverse transport coefficient is less constrained due to the influence of the underlying mechanism for heavy-quark medium interaction. Additional constraints on transport models such as energy loss fluctuation and transverse-momentum broadening can further reduce theoretical uncertainties in the extracted transport coefficients.

\end{abstract}

\maketitle

%%%%%%%%%%%%%%%%%%%%%%%%%%%%%%%%%%%%%%%%%%%%%%%%%%%%%%%%%%%%%%%%%%%%%

\section{Introduction}
\label{sec:Introduction}

Hard probes such as large transverse-momentum ($p_\mathrm{T}$) jets and heavy-flavor (HF) hadrons play an essential role in the study of the properties of the quark-gluon plasma (QGP) created in high-energy heavy-ion collisions. The  energy-momentum scale typically involved with these hard probes is large enough to enable perturbative-QCD (pQCD) calculations of their initial production rate and, at high $p_\mathrm{T}$, of the medium modification of the final spectra and correlations. They can therefore provide important information  about the hot QCD medium probed by these particles. Because of their large mass the thermal production of heavy quarks is negligible in the QGP within the range of temperatures that can be reached in heavy-ion collisions at the Relativistic Heavy Collider (RHIC) and the Large Hadron Collider (LHC). Therefore heavy-quark (HQ) physics utilizes the modification of their spectra caused by the  interactions with the light quarks and gluons during their propagation in a dynamically evolving QCD medium.

At high momentum, the propagation of heavy quarks is similar to that of energetic light quarks and gluons. Their interactions with the medium can be described by scattering with medium partons. Perturbative-QCD calculations \cite{Gyulassy:1993hr,Baier:1996sk,Zakharov:1996fv,Gyulassy:2000er,Wiedemann:2000za,Wang:2001ifa,Arnold:2002ja} show that the energy loss experienced by  high-energy partons is dominated by induced gluon radiation that leads to a suppression of final hadrons with large $p_\mathrm{T}$, known as jet quenching \cite{Gyulassy:1990ye,Wang:1991xy}. The parton energy loss and the suppression factor for final leading high-$p_\mathrm{T}$ hadrons is determined by a jet transport coefficient, $\hat q(E)$ \cite{Baier:1996sk}, which is essentially the average transverse momentum broadening squared per unit length of propagation  of an energetic parton with an energy $E$.  Such a jet transport coefficient encodes the coupling between the jet parton and the medium, as well as its energy density, at the energy and momentum scale of typical scatterings\cite{CasalderreySolana:2007sw,Majumder:2007hx,Liang:2008vz}. It is therefore an important property of the QGP medium as probed by propagating energetic partons. In the limit of the jet parton energy approaching that of a thermal parton $E\sim T$, the jet transport coefficient has been related to the shear viscosity \cite{Majumder:2007zh}, $\eta/s\approx 1.25 T^3/\hat q$,  and hence to the bulk properties of the medium characterizing the coupling among medium partons. 

The large mass of heavy quarks has several implications in this context. It suppresses small-angle gluon radiation leading to smaller radiated energy loss as compared to light quarks and gluons \cite{Dokshitzer:2001zm,Zhang:2003yn,Armesto:2003jh,Aichelin:2013mra}. At low momentum, elastic scatterings become dominant. Since thermal pair production and annihilation processes are negligible, HQ propagation through the hot medium can be described as a diffusion process akin to Brownian motion. The large mass also slows down the equilibration rate of heavy quarks in the medium relative to their light counterparts. The non-equilibrated heavy quarks in the final state can therefore provide information on their interaction with medium throughout their propagation in the QGP medium. The spatial diffusion constant, $D_s$, characterizes the low-momentum interaction strength of heavy quarks in the medium, and has also been related to the shear viscosity of the medium, $D_s(2\pi T) \sim \eta/s$\cite{Rapp:2009my}.  It encodes the $p_\mathrm{T}$ broadening of the heavy quark,  while the drag coefficient $A$ describes the longitudinal-momentum or energy loss in the diffusion process. In this way HQ transport yields valuable information on the coupling strength and properties of the interaction in the QGP \cite{Svetitsky:1987gq,vanHees:2004gq,Moore:2004tg}.

Since the first observation of jet quenching at RHIC in 2001 \cite{Adcox:2001jp,Adams:2003kv}, experimental studies of hard probes at both RHIC and the LHC have generated an enormous amount of precision data on the medium modification of high-$p_\mathrm{T}$ light- and heavy-flavor hadrons \cite{Adcox:2004mh,Adams:2005dq,Wang:2004dn,Jacobs:2004qv,Majumder:2010qh,Muller:2012zq,Qin:2015srf}. A systematic study of the experimental data on the suppression of high-$p_\mathrm{T}$ light hadrons at both RHIC and the LHC by the JET Collaboration \cite{Burke:2013yra} has provided the most precise extraction of the jet transport coefficient $\hat q$ to date. The approach adopted by the JET Collaboration is to have a comparative study of high-$p_\mathrm{T}$ hadron suppression of the different theoretical models with the same evolution of the underlying bulk medium, given by the most advanced hydrodynamic models that are constrained by experimental bulk hadron spectra. Such an approach has considerably reduced the theoretical uncertainties in the extraction of the jet transport coefficient. 

The study of the experimental  heavy-hadron spectra and the extraction of HQ transport coefficients is in a similar situation as the light quark/hadron sector before the study by the JET Collaboration. Many phenomenological studies on heavy-hadron spectra with different theoretical models have been carried out \cite{Armesto:2005iq,vanHees:2005wb,Wicks:2007am,Akamatsu:2008ge,Das:2010tj,Alberico:2011zy,He:2011qa,Gossiaux:2012cv,Uphoff:2012gb,Uphoff:2014hza,Song:2015sfa,Song:2015ykw,Cao:2013ita,Djordjevic:2013xoa,Kang:2016ofv,Cao:2016gvr,Cao:2017hhk}. The values of the extracted HQ transport coefficients in these models vary by up to factor of $\sim$5 at high momenta\cite{Xu:2017obm}. The extracted HQ diffusion constant at zero momentum has an uncertainty of about a factor of 3 \cite{Prino:2016cni}. These large variations indicate the need for a systematic and comparative study of the existing models in order to narrow down the theoretical uncertainties in future phenomenological studies. 

In this paper, we report on a coordinated effort  under the auspices of the JET Collaboration to systematically examine six different transport models for charm meson production in heavy-ion collisions and compare their results on the final charm meson suppression and the extracted HQ transport coefficients.  The six commonly used models include the Duke model with Langevin approach \cite{Cao:2013ita,Cao:2015hia,Xu:2015iha}, the Linear Boltzmann Transport (LBT) model \cite{Li:2010ts,Wang:2013cia,He:2015pra,Cao:2016gvr,Cao:2017hhk,Chen:2017zte,Luo:2018pto} by the Central China Normal University (CCNU)  and the Lawrence Berkeley National Laboratory (LBL)  group, the EPOS2+MC@sHQ model \cite{Gossiaux:2009mk,Nahrgang:2013saa,Nahrgang:2013xaa} with a modified pQCD approach, the Texas A{\&}M University (TAMU) model \cite{He:2011qa}  based on the T-matrix approach for non-perturbative HQ interaction with the medium, the Catania quasi-particle Boltzmann approach \cite{Scardina:2017ipo,Das:2015ana} and the Frankfurt Parton Hadron String Dynamics (PHSD) model \cite{Cassing:2008sv,Cassing:2009vt,Bratkovskaya:2011wp,Cassing:2008nn}. We dissect and identify the causes of the variation in extracted HQ transport coefficients from these six transport models by systematically comparing the results with different tunes of each model and in different setups of a static brick QGP medium.  The purpose of this work is to scrutinize the origin of the differences in the models rather than to make a critical evaluation of different models. This systematic study will help to reduce the theoretical and modeling uncertainties in future efforts toward a precision extraction of HQ transport coefficients in the QGP formed in high-energy heavy-ion collisions.

The remainder of this paper is organized as follows. We start with a brief description of the six HQ transport models in Sec.~\ref{sec:models}. In Sec.~\ref{sec:coefficient} we compare the results of the drag $A$ and the jet transport coefficient $\hat q$ calculated from each model in a common basic setup within a pQCD-only treatment of elastic scattering at a fixed value of the strong coupling constant.  We then compare transport coefficients calculated from the six models with both, default parameters and parameters tuned to fit the experimental data on charm $D$ meson suppression in central Pb+Pb collisions at LHC in Sec.~\ref{sec:status}.  In order to eliminate differences in the modeling of the bulk medium evolution and the HQ hadronization, we calculate and compare HQ transport coefficients, in Sec.~\ref{sec:brickProblem},  with an implementation of each model that is tuned to give a fixed value of the HQ suppression factor at a given transverse momentum in a static QGP medium ``brick".  We summarize our study and discuss its implications for future extraction of HQ transport coefficients in heavy-ion collisions in Sec.~\ref{sec:summary}.

This project was proposed and carried out around the same time as a similar but more extended effort within the EMMI Rapid Reaction Task Force frame. The report of this effort has been published in Ref.~\cite{Rapp:2018qla}.

\section{Transport models of heavy quarks}
\label{sec:models}

Various transport models have been developed to investigate the medium modification of heavy flavor production in heavy-ion collisions. In this work, we will employ  six different  model approaches to HQ transport in the QGP and the formation of final charm mesons in heavy-ion collisions. All of these models have been used to extract the heavy quark diffusion coefficient through comparisons to experimental data on charmed meson spectra in heavy-ion collisions for  $p_\mathrm{T}$ up to  5-10 GeV/$c$. We will systematically compare the results on the charm-meson suppression and  the extracted HQ transport coefficients  from these models in an expanded $p_\mathrm{T}$ range (up to 30 GeV/$c$). In this section we  briefly review each model.

%%%%%%%%%%%%%%%%%%%%%%%%%%%%%%%%%%%%%%%%%%%%%%%%%%%%%%%%%%%%%%%%

\subsection{Duke approach}
\label{subsec:Duke}

The model for the space-time evolution of heavy quarks in heavy-ion collisions of the Duke QCD group is based on an improved Langevin approach~\cite{Cao:2013ita,Cao:2015hia,Xu:2015iha}, in which the HQ transport coefficients are extracted via a systematical model-to-data comparison with the Bayesian method~\cite{Bernhard:2016tnd,Xu:2017obm}.

The initial momentum distribution of heavy quarks is calculated using the Fixed Order + Next-to-Leading-Order (FONLL) framework~\cite{Cacciari:1998it, Cacciari:2001td}. To take into account shadowing effects in pA and AA collisions we employ the EPS09 next-to-leading-order (NLO) nuclear parton distribution functions (PDFs)~\cite{Eskola:2009uj},  to calculate the modified HQ initial momentum distribution, from which the initial momenta of heavy quarks are sampled in a Monte Carlo method. The initial position of heavy quarks is generated consistently with the initial condition for the QGP medium by the parametric initial condition model \trento~\cite{Moreland:2014oya,Ke:2016jrd}. At the soft medium thermalization time ($\tau_0=0.6$ fm/c), \trento maps the entropy density $s(x,y)|_{\tau_0}$ to the nucleon thickness function $T_A, T_B$ by evaluating a generalized ansatz at a specific case $s(x,y)|_{\tau_0} \propto  \sqrt{T_AT_B}$. The HQ initial position is then sampled based on the binary collision scaling and is determined by thickness function $\hat{T}_{AB} = T_AT_B$. In this way, the HQ initial position can be related to the spatial distribution of initial soft medium production.

After their production, heavy quarks propagate in the QGP medium and experience energy loss through the interaction with a thermal medium of massless partons. At low momenta, HQ propagation in the QGP medium is treated as a Brownian motion with the assumption that the momentum transfer between the heavy quarks and the medium constituents is small compared to the HQ mass. For the intermediate- and high-momentum region, the radiative energy loss of heavy quarks becomes important;  a recoil force is introduced in order to account for this component. The improved Langevin equation that describes HQ motion is therefore expressed as
\begin{equation}
\frac{d\vec{p}}{dt} = -\eta_D(p) \vec{p} + \vec{\xi} + \vec{f}_g \ .
\end{equation}
The first two terms on the right hand side of the equation are the drag and random thermal forces inherited from the standard Langevin equation. With the requirement that the HQ distribution eventually reaches equilibrium in a thermal medium, a simplified form of the Einstein relation, $\eta_D(p)=\hat{q}/(4TE)$, is used, where $\hat{q}$ is the HQ jet transport coefficient, $T$ is the medium temperature and $E$ is the HQ energy. Assuming a Gaussian-shaped white noise, the thermal random force satisfies the relation $\left<\xi_i(t)\xi_j(t')\right> =  \hat{q} \delta_{ij} \delta(t-t')/2$, which indicates no correlation between thermal forces at different times.

For the radiative energy loss, the Duke approach uses the medium-induced gluon spectra from the higher-twist formalism~\cite{Guo:2000nz,Zhang:2003wk} to calculate the probability of gluon emission from heavy quarks:
\begin{equation}
\frac{dN_g}{dx dk_\perp^2 dt}=\frac{2\alpha_s C_A \hat{q} P(x)k_\perp^4}{\pi \left({k_\perp^2+x^2 M^2}\right)^4} \, {\sin}^2\left(\frac{t-t_i}{2\tau_f}\right),
\label{eq:gluondistribution}
\end{equation}
where $x$ is the fractional energy carried by the emitted gluon, $k_{\perp}$ is the gluon transverse momentum,
$\alpha_s$ is the strong coupling constant, $C_A=N_c$ is the gluon color factor, $P(x)$ is the splitting function, 
and $\hat{q}$ is the jet parton transport parameter. The mass effect on gluon emission from the heavy quark is included in Eq.~(\ref{eq:gluondistribution}). In addition, $t_i$ denotes an ``initial time", or the production time of the parent parton from which the gluon is emitted, and $\tau_f={2Ex(1-x)}/{(k_\perp^2+x^2M^2)}$ is the formation time of the radiated gluon. The recoil force acting on heavy quarks is hence $\vec{f}_g = -d\vec{p}_g/dt$ where $\vec{p}_g$ is the emitted gluon momentum.

Under this construction, the drag force, the thermal random force and the recoil force are dependent on the HQ jet transport coefficient or transport parameter $\hat{q}$, which characterizes the interaction strength between the heavy quarks and the medium. In this study, the HQ transport parameter $\hat{q}$ is related to its spatial coefficient via $\hat{q} = 8\pi T^3 / (D_s 2\pi T)$. Note that although this relation is from the fluctuation-dissipation theorem for heavy quark diffusion near zero momentum, where $D_s$ is conventionally defined, we extend it to finite momentum for parametrizing $\hat{q}$ via~\cite{Xu:2017obm}:
\begin{equation}
\label{eqn:D2piT}
\begin{split}
D_s2\pi T(T, \bm{p}) & = \frac{1}{1 + (\gamma^2 p)^2} (D_s2\pi T) ^{\mathrm{soft}} \\ & + \frac{(\gamma^2 p)^2}{1 + (\gamma^2 p)^2} (D_s2\pi T)^{\mathrm{pQCD}}.
\end{split}
\end{equation}
Here $(D_s2\pi T)^{\mathrm{soft}} = \alpha \left[1+\beta  (T/T_c-1)\right]$ is the soft component which accounts for the non-perturbative effects, and $(D_s2\pi T)^{\mathrm{pQCD}}$ is calculated with pQCD approach at the leading-order with a fixed coupling constant $\alpha_\mathrm{s}=0.3$. The 3 parameters, $\alpha$, $\beta$ and $\gamma$, are determined ($\alpha=1.89$, $\beta=1.59$ and $\gamma=0.26$) using the Bayesian method in comparing the model calculation to experimental data of the heavy-meson nuclear modification factor $R_\mathrm{AA}$ and elliptic flow $v_2$ at RHIC and the LHC.  

The evolution of the QGP medium is simulated by a (2+1)-dimensional event-by-event viscous hydrodynamical model VISHNEW~\cite{Song:2007ux,Shen:2014vra,Heinz:2015arc}. All parameters of the hydrodynamic model, including the temperature-dependent shear and bulk viscosities, have been calibrated to soft hadron spectra using a Bayesian analysis~\cite{Bernhard:2016tnd}.

Once the temperature drops below the critical temperature ($T_\mathrm{c}=154$ MeV), heavy quarks hadronize into heavy mesons through a hybrid model of fragmentation and recombination. The momentum spectra of the heavy mesons that are formed through the recombination process are determined by the Wigner function~\cite{Cao:2013ita,Cao:2015hia},
\begin{equation}
\frac{dN_M}{d^3p_M} = \int d^3 p_1 d^3 p_2 \frac{dN_Q}{d^3p_Q} \frac{dN_q}{d^3 p_q} f^W_M(\vec{p}_Q, \vec{p}_q) \delta(\vec{p}_M - \vec{p}_Q - \vec{p}_q),
\end{equation}
where $\vec{p}_Q$ and $\vec{p}_q$ are the heavy- and light-quark momenta that constitute the heavy meson, $f^W_M(\vec{p}_Q, \vec{p}_q)$ is the Wigner function calculated by overlapping the initial state partons and final meson wavefunction. For heavy quarks that do not combine with light quarks, fragmentation process via \textsc{Pythia} take place.

Below $T_\mathrm{c}$, the hadronic interaction between heavy and light flavor hadrons is then simulated within the Ultra-relativistic Quantum Molecular Dynamics (UrQMD) model by solving the Boltzmann equation for all the particles in the system. The system continues evolves until the hadron gas is so dilute that all the interaction ceases. 

%%%%%%%%%%%%%%%%%%%%%%%%%%%%%%%%%%%%%%%%%%%%%%%%%%%%%%%%%%%%%%%%

\subsection{CCNU-LBNL approach}
\label{subsec:Berkeley}

A Linear Boltzmann Transport (LBT) model has been developed by the CCNU-LBNL group to describe the jet shower parton evolution inside the QGP~\cite{Li:2010ts,Wang:2013cia,He:2015pra,Cao:2016gvr,Cao:2017hhk,Chen:2017zte,Luo:2018pto}. In the absence of a mean field, the evolution of the phase space distribution of a hard parton ``1" (a heavy quark or an energetic light-flavor parton) with $p_1^\mu = (E_1, \vec{p}_1)$ is described with the Boltzmann equation
\begin{equation}
  \label{eq:boltzmann1}
  p_1^\mu \partial_\mu f_1(x_1,p_1)=E_1 (\mathcal{C}_\mathrm{el}+\mathcal{C}_\mathrm{inel}) \ ,
\end{equation}
in which $\mathcal{C}_\mathrm{el}$ and $\mathcal{C}_\mathrm{inel}$ are collision integrals for elastic and inelastic scatterings.

For elastic scattering, the collision term $\mathcal{C}_\mathrm{el}$ is evaluated with the leading-order matrix elements for all possible ``$12\rightarrow34$" scattering processes between the jet parton ``1" and a massless thermal parton ``2" from the medium background. To regulate the collinear ($u,t\rightarrow 0$) divergence of the matrix element, a factor $S_2(s,t,u)=\theta(s\ge2\mu_\mathrm{D}^2)\theta(-s+\mu_\mathrm{D}^2\le t\le -\mu_\mathrm{D}^2)$ is imposed where $\mu_\mathrm{D}^2=g^2T^2(N_c+N_f/2)/3$ is the Debye screening mass. The elastic scattering rate of parton ``1" can then be evaluated as
\begin{align}
 \label{eq:rate2}
 \Gamma_\mathrm{el}&=\sum_{2,3,4}\frac{\gamma_2}{2E_1}\int \frac{d^3 p_2}{(2\pi)^3 2E_2}\int\frac{d^3 p_3}{(2\pi)^3 2E_3}\int\frac{d^3 p_4}{(2\pi)^3 2E_4}\nonumber\\
&\times f_2(\vec{p}_2)\left[1\pm f_3(\vec{p}_3) \right]\left[1\pm f_4(\vec{p}_4)\right] S_2(s,t,u)\nonumber\\
&\times (2\pi)^4\delta^{(4)}(p_1+p_2-p_3-p_4)|\mathcal{M_\mathrm{12\rightarrow34}}|^2,
\end{align}
in which $\gamma_2$ is the spin-color degeneracy of thermal parton ``2." The probability of elastic scattering of parton ``1" in each small time step $\Delta t$ is then $P_\mathrm{el}=\exp (-\Gamma_\mathrm{el}\Delta t)$. 

For inelastic scattering, or the medium-induced gluon radiation process, the LBT model by the CCNU-LBNL group employs the same higher-twist energy loss formalism \cite{Guo:2000nz,Majumder:2009ge,Zhang:2003wk} in Eq.~(\ref{eq:gluondistribution}) as in the Duke approach. The jet transport parameter $\hat{q}$ due to elastic scattering is evaluated with Eq.~(\ref{eq:rate2}) weighted by the transverse momentum broadening of parton ``1." The average number of emitted gluons from a hard parton in each time step $\Delta t$ can be evaluated as \cite{Cao:2013ita,Cao:2015hia,Cao:2016gvr},
\begin{equation}
 \label{eq:gluonnumber}
 \langle N_g \rangle(E,T,t,\Delta t) = \Delta t \int dxdk_\perp^2 \frac{dN_g}{dx dk_\perp^2 dt},
\end{equation}
where a lower cut-off $x_\mathrm{min}=\mu_D/E$ is imposed for the energy of the emitted gluon to avoid possible divergences as $x\rightarrow 0$. Multiple gluon emission is allowed in each time step. Different emitted gluons are assumed independent of each other, and thus their number $n$ obeys a Poisson distribution  
\begin{eqnarray}
\label{eq:possion}
P(n)=\frac{\langle N_g\rangle^n}{n!}e^{-\langle N_g\rangle^n}
\end{eqnarray}
with the mean $\langle N_g \rangle$. The probability for the inelastic scattering process is then $P_\mathrm{inel}=1-e^{-\langle N_g \rangle}$. Note that for the $g\rightarrow gg$ process, $\langle N_g\rangle/2$ is taken as the mean instead to avoid double counting.

To combine elastic and inelastic processes, the total scattering probability is divided into two parts: pure elastic scattering with probability $P_\mathrm{el}(1-P_\mathrm{inel})$ and inelastic scattering with probability $P_\mathrm{inel}$. The total scattering probability is then $P_\mathrm{tot}=P_\mathrm{el}+P_\mathrm{inel}-P_\mathrm{el}\cdot P_\mathrm{inel}$. Based on these probabilities, the Monte Carlo method can be implemented to determine whether a given jet parton is scattered inside the thermal medium and whether the scattering is purely elastic or inelastic. With a selected scattering channel, the energies and momenta of the outgoing partons are sampled based on the corresponding differential spectra given by Eq.~(\ref{eq:rate2}) and (\ref{eq:gluondistribution}). 

To study the evolution of heavy quarks in heavy-ion collisions~\cite{Cao:2016gvr,Cao:2017hhk},  the momentum space distribution of heavy quarks is initialized with the leading-order perturbative QCD (LO pQCD) calculation~\cite{Combridge:1978kx} that includes the pair production ($gg\rightarrow Q\bar{Q}$ and $q\bar{q}\rightarrow Q\bar{Q}$) and the flavor excitation processes ($gQ\rightarrow gQ$ and $g\bar{Q}\rightarrow g\bar{Q}$). The CTEQ parametrizations \cite{Lai:1999wy} and the EPS09 parametrizations~\cite{Eskola:2009uj} of nuclear shadowing are used for the parton distribution functions inside nuclei. The spatial distribution of the HQ production vertices in nucleus-nucleus collisions is sampled using the Monte-Carlo Glauber model. The QGP medium is simulated via a (2+1)-dimensional viscous hydrodynamic model VISHNEW~\cite{Song:2007fn,Song:2007ux,Qiu:2011hf}, in which the Monte-Carlo Glauber model is used to determine the initial entropy density distribution of the hydrodynamic profiles. The starting time of the QGP evolution is set as $\tau_0=0.6$~fm and the shear-viscosity-to-entropy-density ratio ($\eta/s$=0.08) is tuned to describe the spectra of soft hadrons emitted from the QGP fireballs for both RHIC and LHC environments. With this setup, the LBT model is coupled to the hydrodynamic medium to simulate the evolution of heavy quarks inside the QGP above a critical temperature (set as $T_\mathrm{c} = 165$~MeV). In the LBT model, the strong coupling constant $\alpha_\mathrm{s}$ is treated as a model parameter.  A momentum-dependent factor,  $K_p=1+A_p e^{-|\vec{p}|^2/2\sigma_p^2}$, is applied to the HQ transport parameter $\hat{q}$ to include non-perturbative effects beyond the perturbative calculation. The related parameters $A_p=5$ and $\sigma_p=5$~GeV are fixed in earlier works~\cite{Cao:2016gvr,Cao:2017hhk}. On the hadronization hypersurface of the QGP,  a hybrid model of fragmentation plus coalescence~\cite{Cao:2013ita,Cao:2015hia,Cao:2016gvr} is applied,  as already described in Sec.~\ref{subsec:Duke}, to convert heavy quarks into heavy-flavor hadrons. The LBT framework treats heavy- and light-flavor parton evolution on the same footing and allows for a simultaneous description of the nuclear modification of both heavy- and light-flavor hadrons at RHIC and the LHC~\cite{Cao:2017hhk}.

%%%%%%%%%%%%%%%%%%%%%%%%%%%%%%%%%%%%%%%%%%%%%%%%%%%%%%%%%%%%%%%%

\subsection{Nantes approach}
\label{subsec:Nantes}

The Nantes approach is a combination of two major computer programs, EPOS2~\cite{Werner:2010aa} and  the heavy-quark Monte Carlo  MC@HQ~\cite{Gossiaux:2009mk}. EPOS2 is an event generator which describes the soft physics of up, down and strange quarks produced in p+p, p+A and A+A collisions at RHIC and LHC energies. It's results compare fairly well with a large body of experimental data.  The expansion of the QGP after its initial formation is described by hydrodynamical equations. Hadrons are produced employing the Cooper-Frye formula at the transition temperature, and the further hadronic interactions are described by UrQMD. 

The MC@HQ part of the program generates heavy quarks with a FONLL distribution~\cite{Cacciari:1998it, Cacciari:2001td} at the interaction points of nucleon-nucleon collisions during the initial stage of EPOS. Heavy quarks propagate through the QGP and experience elastic~\cite{Gossiaux:2009mk} and radiative collisions~\cite{Aichelin:2013mra,Gossiaux:2010yx} with the plasma constituents (assumed to be massless). In inelastic collisions a gluon is emitted in addition to the particles in the entrance state. The Landau-Pomeranchuck-Migdal effect for radiated gluons is also taken into account, which implies that radiated gluons need time to be considered as independent particles. 

To perform each collision the momentum of the colliding parton from the medium ($q$, $g$) is sampled randomly from the local thermal distribution in the hydrodynamic cell. This parton collides with the heavy quark according to leading-order pQCD cross sections. The elastic cross section differs from the simple pQCD cross section by having a running coupling constant $\alpha (q^2)$ and a modified propagator. Instead of a propagator $\propto (t-\mu_\mathrm{D}^2  )^{-1}$ we use  $\propto (t-\kappa \mu_\mathrm{D}^2  )^{-1}$ where $\kappa$ is determined by the requirement that the energy loss is independent from the intermediate scale which separates the low-momentum transfer dominated by hard thermal loops (HTLs) from the Born diagram which describes the cross section for high-momentum transfer following the procedure given by Braaten and Thoma for QED~\cite{Braaten:1991jj}.

When the QGP in EPOS hadronizes, low-momentum heavy quarks coalesce with a light ($u$, $d$) quarks from the hydrodynamic cell where the heavy quark is localized. For heavy quarks with high momenta the hadronization is obtained by fragmentation based on the Braaten-Cheung-Fleming-Yuan (BCFY) framework in the FONLL approach ~\cite{Cacciari:1998it, Cacciari:2001td}.
After hadronization, UrQMD is used for the final hadronic interactions of $D$ mesons with other hadrons in the medium. EPOS2+MC@HQ has not only been used to compare the results with experimental data on heavy-hadron spectra but also, among others, to study correlations between a heavy quark and antiquark~\cite{Nahrgang:2013saa}, higher order flow components~\cite{Nahrgang:2014vza} and the influence of the existence of hadronic bound states beyond $T_\mathrm{c}$~\cite{Nahrgang:2013xaa}.

%%%%%%%%%%%%%%%%%%%%%%%%%%%%%%%%%%%%%%%%%%%%%%%%%%%%%%%%%%%%%%%%

\subsection{TAMU approach}
\label{subsec:TAMU}

The transport approach for open heavy-flavor (HF) particles developed at Texas A{\&}M University (TAMU)~\cite{He:2011qa} is based on a non-perturbative treatment suitable for a strongly coupled system for both the macroscopic bulk medium evolution and the microscopic HF interactions therein. The former is realized through 2+1 dimensional ideal hydrodynamic simulations of heavy-ion collisions~\cite{He:2011zx} (based on the original AZHYDRO code~\cite{Kolb:2003dz}), carefully tuned to the measured spectra and elliptic flow of bulk hadron production, while the latter are evaluated within a $T$-matrix approach for HQ interactions in the QGP~\cite{vanHees:2007me,Riek:2010fk,Huggins:2012dj} and heavy-meson interactions in hadronic matter~\cite{He:2011yi}. The transition from quark to hadron degrees of freedom in the HF transport is realized within the resonance recombination model (RRM)~\cite{Ravagli:2007xx} which seamlessly converts heavy-light resonant states generated through the $T$-matrix in the QGP into $D$-mesons as the 
transition temperature is approached from above. 

The interactions of heavy quarks with thermal partons (up, down, strange quarks and gluons with thermal masses $gT/\sqrt{3}$) in the QGP are calculated from a thermodynamic $T$-matrix approach~\cite{Mannarelli:2005pz,vanHees:2007me,Riek:2010fk,Huggins:2012dj}. It is 
characterized by an in-medium two-body scattering equation, 
\begin{equation}
T_{l,a} = V_{l,a} + \frac{2}{\pi}\int_0^\infty k^2 dk  V_{l,a} G_{2} T_{l,a} \ , 
\label{Tmat}
\end{equation}
which includes all possible color channels (e.g., $a=1,8$ for $Q\bar q$ and $a=3,6$ for $Qq$), isospin combinations and the two leading partial waves ($l=S,P$); HQ spin symmetry is assumed implying a degeneracy between $S$=0 and $S$=1 states. The intermediate in-medium heavy-light two-particle propagator, $G_2$,  includes single-parton selfenergies. The key input quantity is the interaction kernel, $V_{l,a}$, which is treated in potential approximation adequate for scattering involving at least on heavy particle (which parametrically suppresses the energy transfer, $q_0 \simeq q^2/2m_Q \ll q$, relative to typical thermal momentum transfers of $q\equiv |\vec q|\simeq T$). This, in turn, enables to employ input potentials extracted from the HQ free energies computed with high-precision lattice-QCD (lQCD). Thus far, we have utilized the pertinent internal energies as potential, $V=U$, as computed in Refs.~\cite{Petreczky:2004pz,Kaczmarek:2007pb}. This assumption is motivated by the fact that entropy effects, which are part of the free energy, $F=U-TS$, should emerge from a calculation of medium effects. In addition, the use of the internal energy generally produces better agreement with lQCD results for HQ susceptibilities, Euclidean quarkonium correlators and the HQ diffusion coefficient~\cite{Riek:2010py}. When applying the potential to heavy-light scattering, we include relativistic corrections which ensure that the correct high-energy perturbative limit is recovered (in Born approximation)~\cite{Riek:2010fk}. An important feature of this framework is that, as the pseudo-critical temperature, $T_{\rm pc}\simeq 170$\,MeV, is approached from above, the screening of the potential weakens thus strengthening the interaction. The resummation of the $T$-matrix in Eq.~(\ref{Tmat}) dynamically generates $D$-meson (or $B$-meson) and diquark resonances in the color-singlet and color-triplet channels, respectively, signaling the onset of hadronization. An important role in this is played by remnants of the confining force as a genuine nonperturbative interaction; it is gradually screened as temperature increases.  

The in-medium heavy-light $T$-matrices are straightforwardly implemented to compute drag and diffusion coefficients for HQ transport~\cite{Svetitsky:1987gq}. In the hadronic phase, we evaluate $D$-meson interactions with surrounding
thermal hadrons ($\pi$, $K$, $\eta$, $\rho$, $\omega$, $K^*$, $N$, $\bar N$, $\Delta$ and $\bar\Delta$) utilizing effective hadronic interactions as available from the literature~\cite{He:2011yi}. Remarkably, the resulting diffusion coefficient close to $T_{\rm pc}$ is quite comparable to the QGP result, suggesting both a continuity and a minimum structure through and around $T_{\rm pc}$. 

The transport coefficients are implemented via relativistic Langevin processes with a hydrodynamic simulation for the medium evolution in heavy-ion collisions, carried out in the local rest frame at the local temperature in a given cell. The $T$-matrix approach accounts for in-medium charm-quark masses defined by the infinite-distance limit of the internal energy, amounting to $m_c$$\simeq$1.8\,GeV close to $T_{\rm pc}$, and slowly decreasing with temperature. This implies that the Fokker-Planck approximation remains accurate until at least $T$=300\,MeV. While the hydro evolution does not include viscosity, it turns out that a suitable tuning of initial conditions (including a compact overlap profile and an initial-flow field), together with lQCD equation of state, enable a reasonable reproduction of $p_\mathrm{T}$ spectra and elliptic flow of light hadrons at RHIC and LHC energies~\cite{He:2011zx}.

The final ingredient is the conversion from quark to hadronic degrees of freedom in the HF transport simulation. This is achieved by applying the resonance recombination model (RRM)~\cite{Ravagli:2007xx} on a hydro-hypersurface at $T_{\rm pc}$ using the ($p$-dependent) $c\to D$ scattering rates from the heavy-light $T$-matrices. The RRM is 4-momentum 
conserving and thus recovers the correct equilibrium limit which has been explicitly verified for $p_\mathrm{T}$ spectra and $v_2$
corresponding to the hydrodynamic flow fields~\cite{He:2011qa}. Heavy quarks which do not recombine are hadronized via FONLL fragmentation~\cite{He:2014cla}, in line with the choice for the initial spectra to recover $D$-meson spectra in $pp$ collisions. In AA collisions, an additional EPS09 shadowing correction is accounted for~\cite{He:2014cla}.

%%%%%%%%%%%%%%%%%%%%%%%%%%%%%%%%%%%%%%%%%%%%%%%%%%%%%%%%%%%%%%%%

\subsection{Catania approach}
\label{subsec:Catania}

In the Quasi-Particle-Boltzmann (QP-BM) approach the propagation of heavy quarks inside the hot QCD medium is described by means of the Boltzmann Equation (BE),
\begin{equation}
\label{eq:HQBoltzmann}
p^{\mu}\partial_{\mu}f_{HQ}(x,p) = C[f_{HQ},f_q,f_g](x,p) \ , 
\end{equation}
similar to the LBT model [Eq.~(\ref{eq:boltzmann1})], where $f_{HQ}(x,p)$ is the single-particle phase-space distribution function for an on-shell heavy quark, while $C$ is the Boltzmann-like collision integral which encodes the dissipative part governing the HQ evolution. The space-time evolution of quark and gluon one-body distribution function $f_g$ and $f_q$ is calculated as in Ref.~\cite{Plumari:2012ep} [see description after Eq.~(\ref{eq:QPMmasses})]. In this work only elastic processes between heavy quarks and bulk partons are considered, i.e. $HQ(p_1)+i(g,q)(p_2)\rightarrow HQ(p_{1}^{\prime})+i(g,q)(p_{2}^{\prime})$. Therefore the collision integral takes the form
\begin{eqnarray}\label{eq:CollisionInt}
{\cal C}[f_{HQ}]&=&\frac{1}{2E_1}  \!\!\!  \sum_{i=g,q} \! \int \!\!\!\frac{d^3p_2}{2E_2(2\pi)^3}  \!\!\! \int  \!\!\! \frac{d^3p_{1}^{\prime}}{2E'_{1}(2\pi)^3} \!\!\!  \int  \!\!\! \frac{d^3p_{2}^{\prime}}{2E'_{2}(2\pi)^3}\nonumber\\
&&\times \left[f_{HQ}(p_{1}^{\prime})f_{i}(p_{2}^{\prime}) - f_{HQ}(p_1) f_{i}( p_2)\right] \nonumber\\
&&\times \frac{1}{\nu_i} |{\cal M}_{HQ+i}(p_1p_2\rightarrow p_1^\prime p_2^\prime)|^2 \nonumber \\
&&  \times(2\pi)^4  \delta^4(p_1 + p_2 - p_{1}^{\prime} - p_{2}^{\prime}).
\end{eqnarray}
In order to solve numerically the BE Eq.~\eqref{eq:HQBoltzmann}, the coordinate space is divided into a three-dimensional (3-D) lattice and the distribution function $f_{HQ}(x,p)$ in each cell   
%of each parton species with $i=HQ,g,q$ in the phase-space   
is sampled according to the test-particle method~\cite{Ferini:2008he}.
A solution of BE is obtained by solving the canonical Hamilton equations for each test particle. 
The key ingredient is represented by the variation of the HQ momentum due to scattering processes with the bulk partons encoded in the collision integral $C$. This kernel is mapped through a stochastic algorithm into a probability of elastic collision~\cite{Xu:2004mz},
\begin{equation}\label{eq:CollProb}
P_{\rm coll}= v_{\rm rel} \sigma_{22} \frac{\Delta t}{\Delta^3x},
\end{equation}
where $v_{\rm rel}$ is the relative velocity between the two scattering particles, $\Delta t$ is the time step of the simulation and $\Delta^3 \vec{x}$ is the volume of the cells. The numerical solution of the Boltzmann equation through the stochastic method converges in the limit of $\Delta t \rightarrow 0$, $\Delta^3x \rightarrow 0$. The total cross section for elastic processes,
\begin{eqnarray}\label{eq:Sigma22}
\sigma_{22} &=& \frac{1}{4 v_{\rm rel} E_1 E_2}  \! \int  \!\!\! \frac{d^3p_{1}^{\prime}}{2E'_{1}(2\pi)^3} \!\!\!  \int  \!\!\! \frac{d^3p_{2}^{\prime}}{2E'_{2}(2\pi)^3} \nonumber \\
&& \times \frac{1}{\nu_i} |{\cal M}_{HQ+i}(p_1p_2\rightarrow p_1^\prime p_2^\prime)|^2 \nonumber \\
&& \times(2\pi)^4  \delta^4(p_1 + p_2 - p_{1}^{\prime} - p_{2}^{\prime}) \ , 
\end{eqnarray}
is calculated from the scattering matrices ${\cal M}_{HQ+i}(p_1p_2\rightarrow p_1^\prime p_2^\prime)$ using the standard leading-order pQCD results. 
 
Within this framework the interaction of heavy quarks with bulk partons is described by means of a Quasi-Particle (QP) model accounting non-perturbative effects in QCD~\cite{Plumari:2011mk}. Light quarks and gluons forming the medium are dressed with thermal masses
\begin{eqnarray}\label{eq:QPMmasses}
 m^2_g(T)&=&\frac{2N_{c}}{N_{c}^2-1}\,g^2(T)\,T^2, \nonumber \\
 m^2_q(T)&=&\frac{1}{N_c}\,g^2(T)\,T^2,
 \end{eqnarray}  
while the $T$-dependence of the strong coupling constant, $g(T)$, for $T>T_c$ follows a logarithmic parametrization,
\begin{equation}\label{eq:QPMcoupling}
g^2(T) = \frac{48\pi^2}{(11N_c-2N_f) \ln \left[ \lambda \left(\frac{T}{T_c} - \frac{T_s}{T_c}\right) \right]^2}.
\end{equation}
which is used also in other models \cite{Peshier:1995ty,Berrehrah:2013mua}. The parameters, $\lambda=2.6$ and $T_s/T_c=0.57$ for a critical temperature $T_c=0.155$ GeV, color number $N_c=3$ and quark flavors $N_f=3$, are fitted to the results on thermodynamics from Wuppertal-Budapest QCD calculations~\cite{Borsanyi:2010cj}.

%BULK EVOLUTION WITHIN QPM
The QGP evolution is described by a modified version of the BE, Eq.~\eqref{eq:HQBoltzmann}, where the interaction between light quarks and gluons is tuned to a fixed value of $\eta/s(T)$ that is realized via locally computing the bulk cross section according to the Chapmann-Enskog approximation~\cite{Plumari:2012ep,Ruggieri:2013ova}. In this way one can gauge the collision integral to the desired $\eta/s(T)$ and simulate the evolution of the fluid in analogy to what is performed within hydrodynamics~\cite{Romatschke:2007mq}.

In realistic simulations charm quarks are distributed in momentum space using a power law fit of the FONLL spectra with shadowing effects parametrized from EPS09 while in coordinate space they are sampled according to the number of binary collisions provided by the standard Glauber model. For a detailed discussion of charm dynamics in QGP within the QP-BM approach and the results for the nuclear modification factor $R_\mathrm{AA}$ and the elliptic flow $v_2$ of $D$ mesons obtained at RHIC and LHC energies one can refer to Refs.~\cite{Scardina:2017ipo,Das:2015ana}.

%BRIEF ON COALESCENCE AND FRAGM 
Finally, the HQ hadronization is performed at the final stage of the transport evolution, and it is based on the hybrid fragmentation plus coalescence approach described in Ref.~\cite{Plumari:2017ntm}. The coalescence model is based on the Wigner formalism and provides a $p_\mathrm{T}$-spectrum of hadrons which can be written as
\begin{eqnarray}\label{eq:Coalspectra}
\frac{dN_{H}}{d^{2}P_{T}\,dy}&=& g_{H} \int \prod^{n}_{i=1} \frac{d^{3}p_{i}}{(2\pi)^{3}E_{i}} p_{i} \cdot d\sigma_{i}  \; f_{q_i}(x_{i}, p_{i})\\ 
&\times& \!\!\! f_{H}(x_{1}...x_{n}, p_{1}...p_{n})\, \delta^{(2)} \!\!\left(\!P_{T}-\sum^{n}_{i=1} p_{T,i}\!\right),\nonumber 
\end{eqnarray}
where $d\sigma_i$ denotes an element of a space-like hypersurface, $f_{q_{i}}$ are the quark (anti-quark) phase-space distribution functions with $n=2,3$ respectively for meson and baryon formation, and $g_H$ is the statistical factor to form a colorless hadron. In particular, for $D$ mesons one has $g_D=1/36$;
$f_{H}(x_{1}...x_{n}, p_{1}...p_{n})$ is the Wigner function which describes the spatial and momentum distribution of quarks inside the hadron. For charmed mesons one can adopt a Gaussian shape with respect to the relative coordinates $x_r=x_1-x_2$ and momentum $p_r=(m_2p_1-m_1p_2)/(m_1+m_2)$,
\begin{equation}\label{eq:Wignergauss}
f_{M}(x_{1}, x_{2}; p_{1}, p_{2})=A_{W}\exp{\Big(-\frac{x_{r}^2}{\sigma_r^2} - p_{r}^2 \sigma_r^2\Big)},
\end{equation}
where $A_W$ is a normalization factor and $\sigma_r$ is a width parameter which depends on the hadron species and can be calculated from the charge radius $<r_{ch}^2>$ according to the quark model~\cite{Hwang:2001th,Albertus:2003sx}. For $D$ mesons this single parameter is fixed in order to have $<r_{ch}^2>=0.184$~fm$^2$ which corresponds to $\sigma_r^{-1}=0.283$~GeV. The coalescence integral in Eq.~\eqref{eq:Coalspectra} is solved numerically within a Monte Carlo method as explained in Ref.~\cite{Greco:2003xt}. 
The fraction of charm quarks which do not undergo to coalescence is indicated as $dN_{\rm frg}/d^2p_\mathrm{T}dy$ and gives rise to the following hadron $p_\mathrm{T}$-spectra 
\begin{equation}\label{eq:Fragspectra}
\frac{dN_{H}}{d^{2}p_\mathrm{T}\,dy}=\sum \int dz \frac{dN_{\rm frg}}{d^{2}p_\mathrm{T}\, dy} \frac{D_{H/c}(z,Q^{2})}{z^{2}} 
\end{equation}
where $z=p_H/p_c$ is the fraction of charm momentum carried away by the leading hadron, while $Q^2=(p_H/2z)^2$ is the momentum scale of the fragmentation process. In Eq.~\eqref{eq:Fragspectra} the Peterson fragmentation function
\begin{equation}\label{eq:Peterson}
D_{H}(z,Q^2) = 1 / \bigg[ {z\left[ 1 - \frac{1}{z} -\frac{\epsilon_c}{1-z}\right]^2} \bigg]
\end{equation}
is employed with $\epsilon_c=0.06$ according to the experimental data of $D$ meson production in $p+p$ collisions~\cite{Scardina:2017ipo}.

In this work, the QPM model is implemented with the Boltzmann approach, while the pQCD model is taken from our earlier Langevin calculation.

%%%%%%%%%%%%%%%%%%%%%%%%%%%%%%%%%%%%%%%%%%%%%%%%%%%%%%%%%%%%%%%%

\subsection{Frankfurt (PHSD) approach}
\label{subsec:Frankfurt}

The Parton-Hadron-String Dynamics (PHSD) transport approach~\cite{Cassing:2008sv,Cassing:2009vt,Bratkovskaya:2011wp,Cassing:2008nn} is a microscopic covariant dynamical model for strongly interacting systems formulated on the basis of Kadanoff-Baym equations \cite{Kadanoff1} for Green's functions in phase-space representation (in first-order gradient expansion beyond the quasi-particle approximation). The approach consistently describes the full evolution of a relativistic heavy-ion collision from the initial hard scatterings and string formation through the dynamical deconfinement phase transition to the strongly-interacting QGP (sQGP), as well as hadronization and the subsequent interactions in the expanding hadronic phase as in the Hadron-String-Dynamics (HSD) transport approach \cite{Cassing:1999es,Cassing:2000bj}.

The transport theoretical description of quarks and gluons in PHSD is based on the Dynamical Quasi-Particle Model (DQPM) for partons that is constructed to reproduce lQCD results for the QGP in thermodynamic equilibrium~\cite{Cassing:2008nn,Berrehrah:2016vzw} on the basis of effective propagators for quarks and gluons.
The DQPM provides the properties of the partons, i.e., masses and widths in their spectral functions as well as the mean fields for gluons and quarks and their effective two-body interactions that are implemented in PHSD \cite{Cassing:2008nn,Linnyk:2015rco}. In equilibrium PHSD reproduces the partonic transport coefficients such as shear and bulk viscosities or the electric conductivity from lQCD calculations as well \cite{Ozvenchuk:2012kh,Linnyk:2015rco}.
The PHSD approach has been applied to p+p, p+A and A+A collisions from lower Schwerionensynchrotron (SIS) to LHC energies and been successful in describing a large number of experimental data including single-particle spectra, collective flow and electromagnetic probes~\cite{Cassing:2009vt,Bratkovskaya:2011wp,Linnyk:2015rco}.

In PHSD the charm and bottom quark pairs are produced through initial hard nucleon-nucleon scattering in relativistic heavy-ion collisions.  The \textsc{Pythia} event generator \cite{Sjostrand:2006za} is employed to produce the HQ pairs whose transverse momentum and rapidity are modified slightly such that they are similar to those from the FONLL calculations \cite{Cacciari:2012ny}. The corrections employed at RHIC and LHC energies can be found in Refs. \cite{Song:2015sfa,Song:2015ykw,Song:2016rzw}. Accordingly, the tuned \textsc{Pythia} generator gives very similar charm and bottom distributions as those from FONLL calculations \cite{Cacciari:2012ny,Cacciari:2005rk}, which provides the input for the initial HQ production.

The produced charm and bottom quarks in hard nucleon-nucleon interactions are hadronized in p+p collisions by emitting soft
gluons, which is denoted by ``fragmentation" (cf. Ref. \cite{Song:2015sfa} for details). The excited $D^* (B^*)$ mesons
first decay into $D (B)+\pi$ or $D (B)+\gamma$, and finally some of the $D$ and $B$ mesons can produce single electrons through semi-leptonic decays~\cite{Agashe:2014kda}. In the case of heavy-ion collisions, the shadowing effect is incorporated in PHSD by employing the EPS09 package from Ref. \cite{Eskola:2009uj}. The details of the implementation are given in Ref. \cite{Song:2015ykw}.

In PHSD the baryon-baryon and baryon-meson collisions at high-energy produce strings. They melt into quarks and antiquarks when the critical energy density ($\sim$ 0.5 GeV/fm$^3$) is reached, with masses determined by the temperature-dependent spectral functions from the DQPM \cite{Cassing:2008nn}, which has been fitted to thermodynamical quantities from lQCD. Massive gluons are formed through flavor-neutral quark and antiquark fusion in line with the DQPM. The heavy quarks and antiquarks produced in early hard collisions interact with the dressed light off-shell partons in the partonic phase. The cross sections for the HQ scattering with massive off-shell partons have been calculated in Ref.~\cite{Berrehrah:2014kba,Berrehrah:2015ywa} including the spectral functions of partons. The elastic scattering of heavy quarks in the QGP is treated in PHSD by including the non-perturbative effects of the sQGP constituents, i.e., the temperature-dependent coupling $g(T/T_c)$ as well as the effective propagators with broad spectral functions (and imaginary parts) from the DQPM~\cite{Cassing:2008nn}. We note that in PHSD HQ interactions
in the QGP, as described by the DQPM charm scattering cross sections, differ substantially from the pQCD scenario, and are constructed such that the spatial diffusion constant for charm quarks $D_s(T)$ is consistent with the lQCD data \cite{Song:2015ykw,Berrehrah:2016vzw}.

The HQ hadronization in heavy-ion collisions is realized via ``dynamical coalescence" in competition to fragmentation. Here ``dynamical coalescence" means that a coalescence partner is decided by Monte Carlo based on coalescence probability in the vicinity of the critical energy density $0.4\le \epsilon \le 0.75$ GeV/fm$^3$ as explained in Ref. \cite{Song:2015ykw}.

After the hadronization of heavy quarks and their subsequent decay into $D, D^*, B$ and $B^*$ mesons, the final mesons follow a realistic description of the hadron-hadron scattering, potentially affected by resonant interactions,  with hadronic states $\pi,K,\bar{K},\eta,N,\bar{N},\Delta,$ and $\bar{\Delta}$ from the expanding bulk medium. Such a description of hadronic interactions has been developed in Refs.~\cite{Tolos:2013kva,Torres-Rincon:2014ffa,Tolos:2013gta} using effective field theory. The resulting cross sections are implemented in PHSD.

%%%%%%%%%%%%%%%%%%%%%%%%%%%%%%%%%%%%%%%%%%%%%%%%%%%%%%%%%%%%%%%%

\section{Heavy-quark transport coefficients with a common basic setup}
\label{sec:coefficient}

Among various transport coefficients that characterize the HQ interaction with a thermal medium, the drag $A$ and the jet transport parameter $\hat{q}$ quantify longitudinal momentum loss and transverse-momentum broadening squared per unit time as the heavy quarks propagate through the medium. In this study, they are defined as
\begin{equation}
A=dp_\mathrm{L}/dt \ , \   
\hat{q}=dp^2_\mathrm{T}/dt \ ,
\end{equation}
Note that when both elastic and inelastic processes are included in a transport approach, $A$ is extracted from the total longitudinal momentum loss unless otherwise specified. On the other hand, the HQ transport parameter $\hat{q}$ is defined by convention through elastic processes only, since it is this elastic part of the transverse-momentum broadening that directly quantifies the rate of the medium-induced gluon emission and thus the inelastic energy loss~\cite{Burke:2013yra} at the lowest order of pQCD.

\begin{figure}
        \centering
        \subfigure{\includegraphics[width=0.45\textwidth]{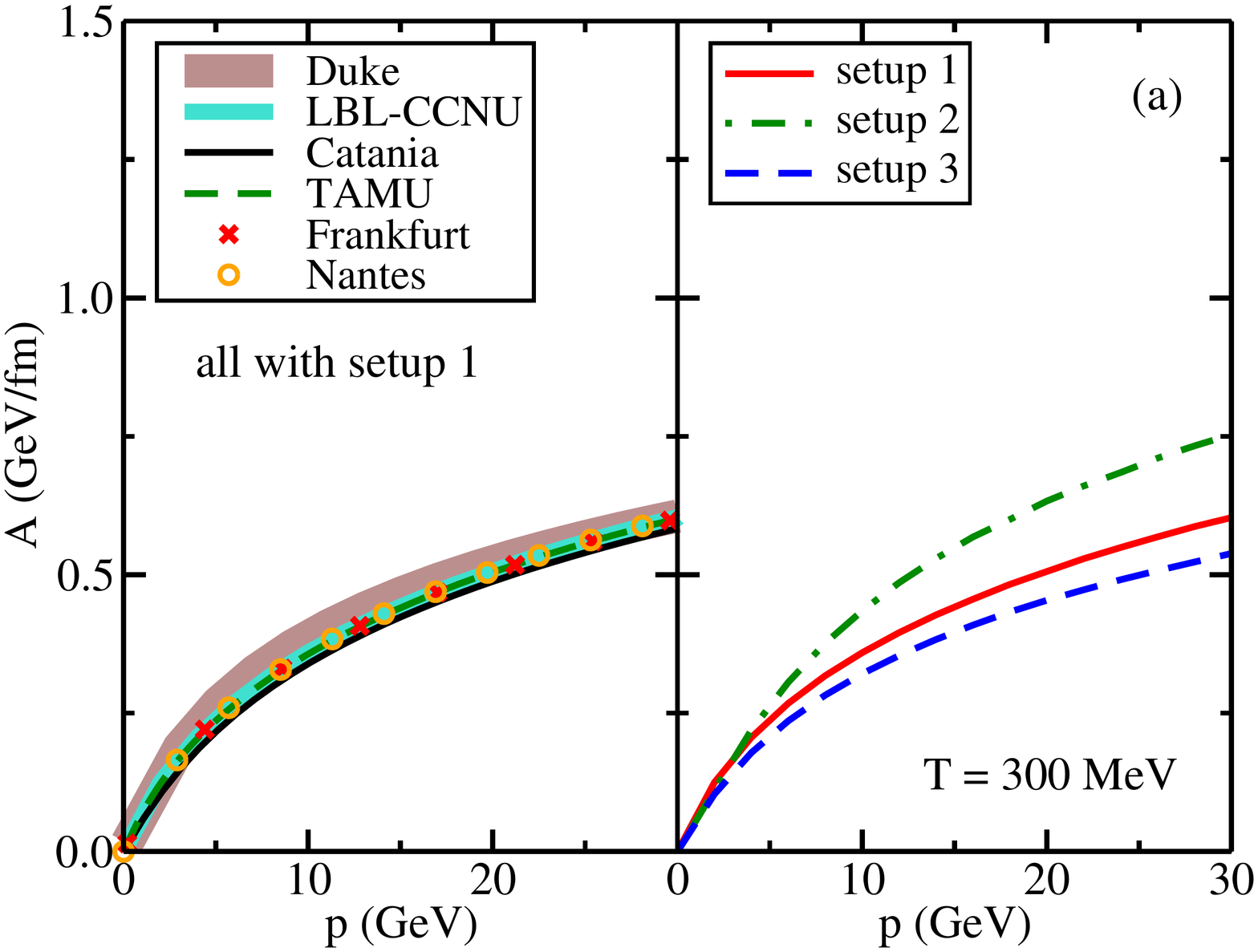}\label{fig:plot-check-drag-p}}\vspace{-20pt}
        \subfigure{\includegraphics[width=0.45\textwidth]{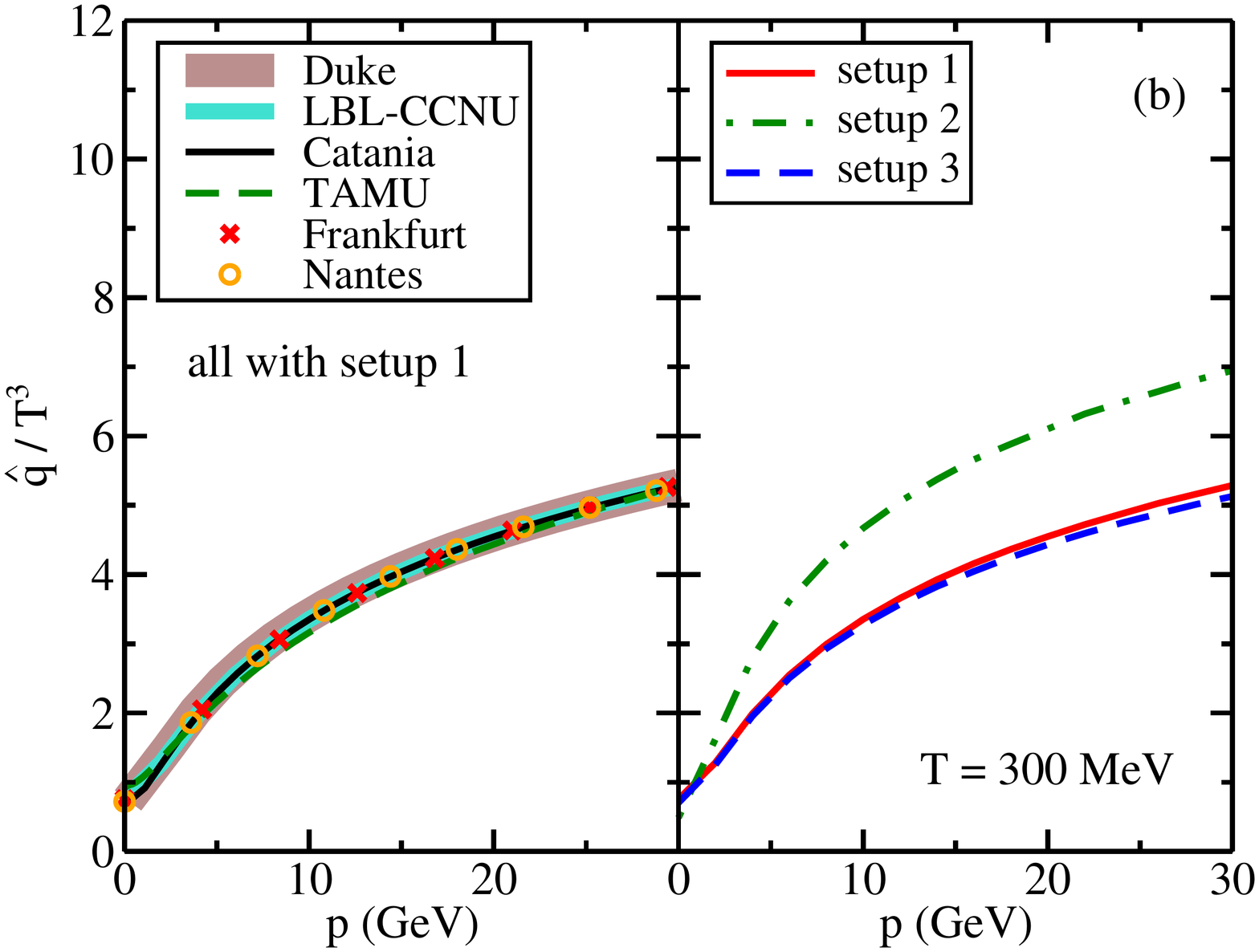}\label{fig:plot-check-qhat-p}}
        \caption{(Color online) Calculations of (a) $A$ and (b) $\hat{q}/T^3$, compared between different groups with a common setup in the left columns, and different setups (within the CCNU-LBNL model) in the right columns (setup 1: $t-\mu^2_\mathrm{D}$ regulator with quantum statistics; setup 2: $t<-\mu^2_\mathrm{D}$ cut-off with quantum statistics;  setup 3: $t-\mu^2_\mathrm{D}$ regulator with classical statistics). }
        \label{fig:check}
\end{figure}

Before systematically extracting $A$ and $\hat{q}$ using the HQ transport  approaches presented in Sec.~\ref{sec:models}, we design a common formalism and compare the calculated $A$ and $\hat{q}$ of the six groups in Fig.~\ref{fig:check}. Only elastic scattering processes between a charm quark ($M_c=1.5$~GeV) and massless thermal partons are taken into account. Both $A$ and $\hat{q}$ are evaluated with the lowest-order pQCD matrix elements,  the strong coupling constant is set as $\alpha_\mathrm{s}=g^2/(4\pi)=0.3$,  the number of thermal quark flavors is set to $n_f=3$, the medium temperature is set to $T=300$~MeV and the Debye screening mass is set to $\mu_\mathrm{D}=gT$.  A Fermi-Dirac/Bose-Einstein distribution is used for the thermal light flavor quark/gluon distribution to take into account the quantum statistics in the initial state. Effects of  Bose enhancement and Pauli blocking in the final state are, however,  not included. To regulate the collinear divergence of the $t$-channel scattering matrix, $1/t\rightarrow1/(t-\mu_\mathrm{D}^2)$ is implemented. As shown in the left columns of Figs.~\ref{fig:plot-check-drag-p} and \ref{fig:plot-check-qhat-p},  consistent values for $A$ and $\hat{q}$ as functions of the HQ momentum are obtained of the six groups with this common setup (denoted as setup 1). This serves as a crucial baseline to verify that the same definitions of transport coefficients are shared by the different groups and are correctly implemented in their calculation.

To study the influence of the different parts of setup 1 on the results, within the CCNU-LBNL model, we first check the result for $A$ and $\hat{q}$ if the infrared regulator in the $t$-channel  $1/t\rightarrow1/(t-\mu_\mathrm{D}^2)$ is replaced by  $t < -\mu_D^2$. This setup is denoted as set up 2. In this case a larger average momentum will be transferred between the heavy quark and the thermal medium, and thus larger values of both $A$ and $\hat{q}$ are expected as seen on the right hand side of Figs.~\ref{fig:check}(a) and \ref{fig:check}(b).  Ignoring the quantum statistics for the thermal parton distribution functions (called setup 3, with $1/(t-\mu_\mathrm{D}^2)$ infrared regulator) leads to slightly smaller values of $A$ and $\hat{q}$ (compared to setup 1). These effects are worth noticing when different detailed implementations are adopted by various model calculations.

\section{Current status of extracting heavy quark transport coefficients}
\label{sec:status}

In most of the models described in Sec.~\ref{sec:models},  model parameters are adjusted to fit the experimental HF hadron spectra in both p+p and A+A collisions at RHIC and LHC. With these model parameters, one can then evaluate or extract HQ transport coefficients. In this section, we will review the model comparisons to experimental data on HF hadron nuclear modification factors, $R_\mathrm{AA}(p_\mathrm{T})$, at both RHIC and LHC energies and the extracted HQ transport coefficients.

\subsection{Model to data comparison}
\label{subsec:data}

\begin{figure}
        \centering
             \subfigure{\includegraphics[width=0.45\textwidth]{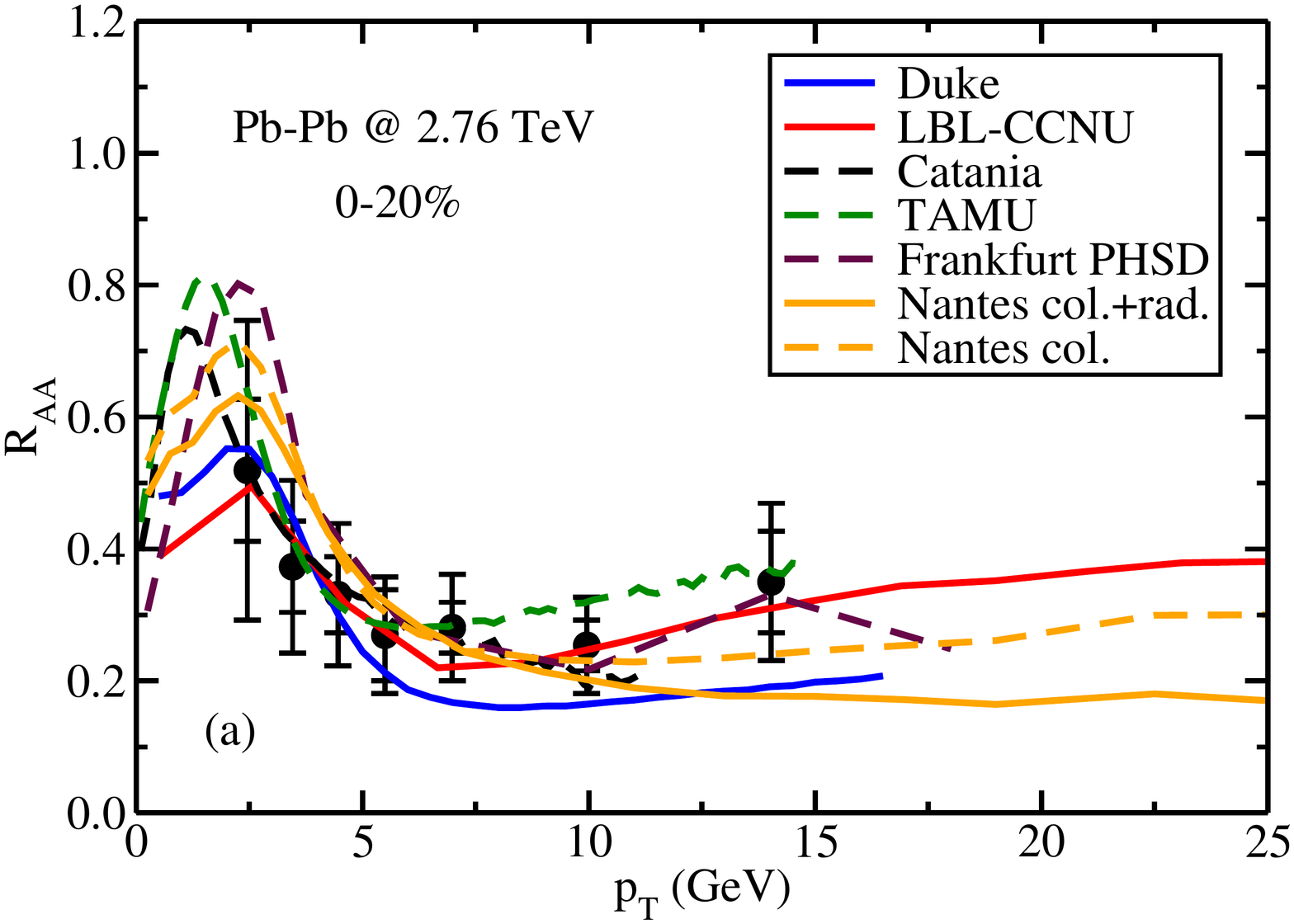}\label{fig:data-LHC}}\vspace{-20pt}
             \subfigure{\includegraphics[width=0.45\textwidth]{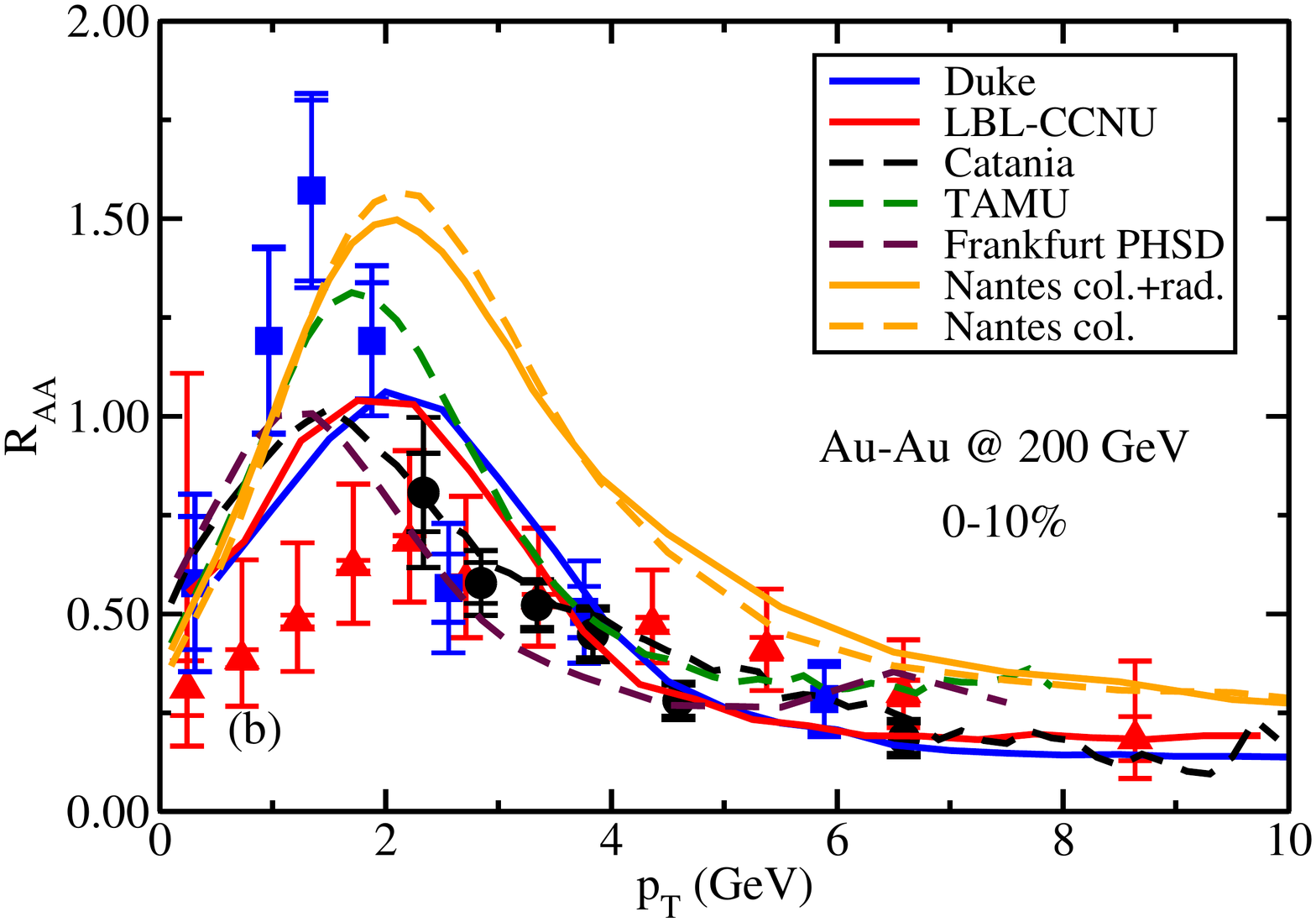}\label{fig:data-RHIC}}
        \caption{Model calculations of the $D$ meson $R_\mathrm{AA}$ (a) with ``tune 1" parameters (see Table~\ref{fig:TC-T}) in central Pb-Pb collisions at 2.76~ATeV and (b) Au-Au collisions at 200~AGeV as compared to experimental data  \cite{ALICE:2012ab, Adamczyk:2014uip,Xie:2016iwq}. }
        \label{fig:data}
\end{figure}

\begin{table}[h]
\begin{tabular}{l | c | c | c | c | c | c}
Models & 2.76~ATeV Pb-Pb & 200 AGeV Au-Au  \\
\hline \hline 
Duke & 0.769 & 2.819   \\ 
\hline
CCNU-LBNL & 0.132 & 1.49 \\ 
\hline
Catania & 0.113 & 1.01 \\
\hline
TAMU & 0.178 & 2.40  \\
\hline
Frankfurt PHSD & 0.637 & 1.59 \\
\hline
Nantes col. + rad. & 0.629 & 17.3 \\
\hline
Nantes col. only & 0.524 & 17.9
\end{tabular}
\caption{Values of $\chi^2$/dof from model to data comparison.}
\label{tab:chi2}
\end{table}

In Fig.~\ref{fig:data}, we summarize the current comparisons between different model calculations, as described in Sec.~\ref{sec:models} and references therein,  and the experimental data on the $D$-meson $R_\mathrm{AA}$ at RHIC and the LHC. The values of the standard deviation $\chi^2$ per degree of freedom (dof) between model calculations and data are presented in Tab.~\ref{tab:chi2}. One observes that with a proper adjustment of model parameters, most transport models are able to describe the experimental data reasonably well. The deviation of the Nantes calculation (EPOS2+MC@sHQ) from data at RHIC results from the bulk matter evolution (EPOS2) that relies on an ideal hydrodynamic model that has not been fine-tuned for heavy-collisions at RHIC. In Fig.~\ref{fig:data-RHIC}, we compare to published data  from Refs.~\cite{Adamczyk:2014uip} (blue squares) and \cite{Xie:2016iwq}(black circles).  Note that the STAR Collaboration released a correction to the published $R_\mathrm{AA}$ data fromthe 2014 Heavy-Flavor-Tracker (HFT) run at the last Quark Matter conference (red triangles in Fig.~\ref{fig:data-RHIC}). The new preliminary results are consistent with the published ones at $p_\mathrm{T}>2$~GeV/c, but the central values of the new results at $p_\mathrm{T}<2$~GeV/c are lower than the published results by about a factor of 2. The publications of the correction are in preparation, as well as plans for new high-precision Au+Au data from future reanalysis.

\subsection{Current extraction of $A$ and $\hat{q}$ }
\label{subsec:currentCoefficient}

\begin{figure}
        \centering
             \subfigure{\includegraphics[width=0.45\textwidth]{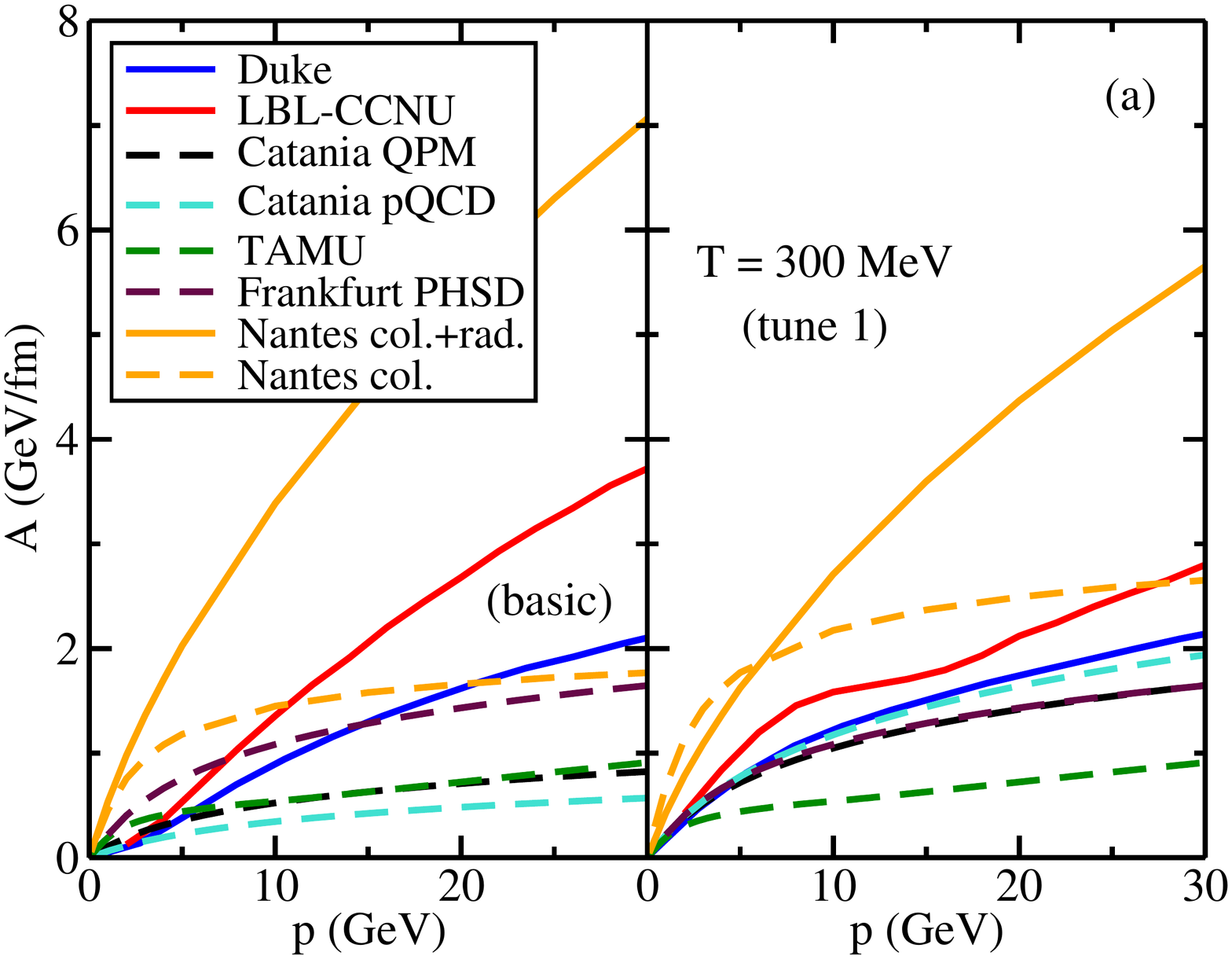}\label{fig:TC-p-drag}}\vspace{-20pt}
             \subfigure{\includegraphics[width=0.45\textwidth]{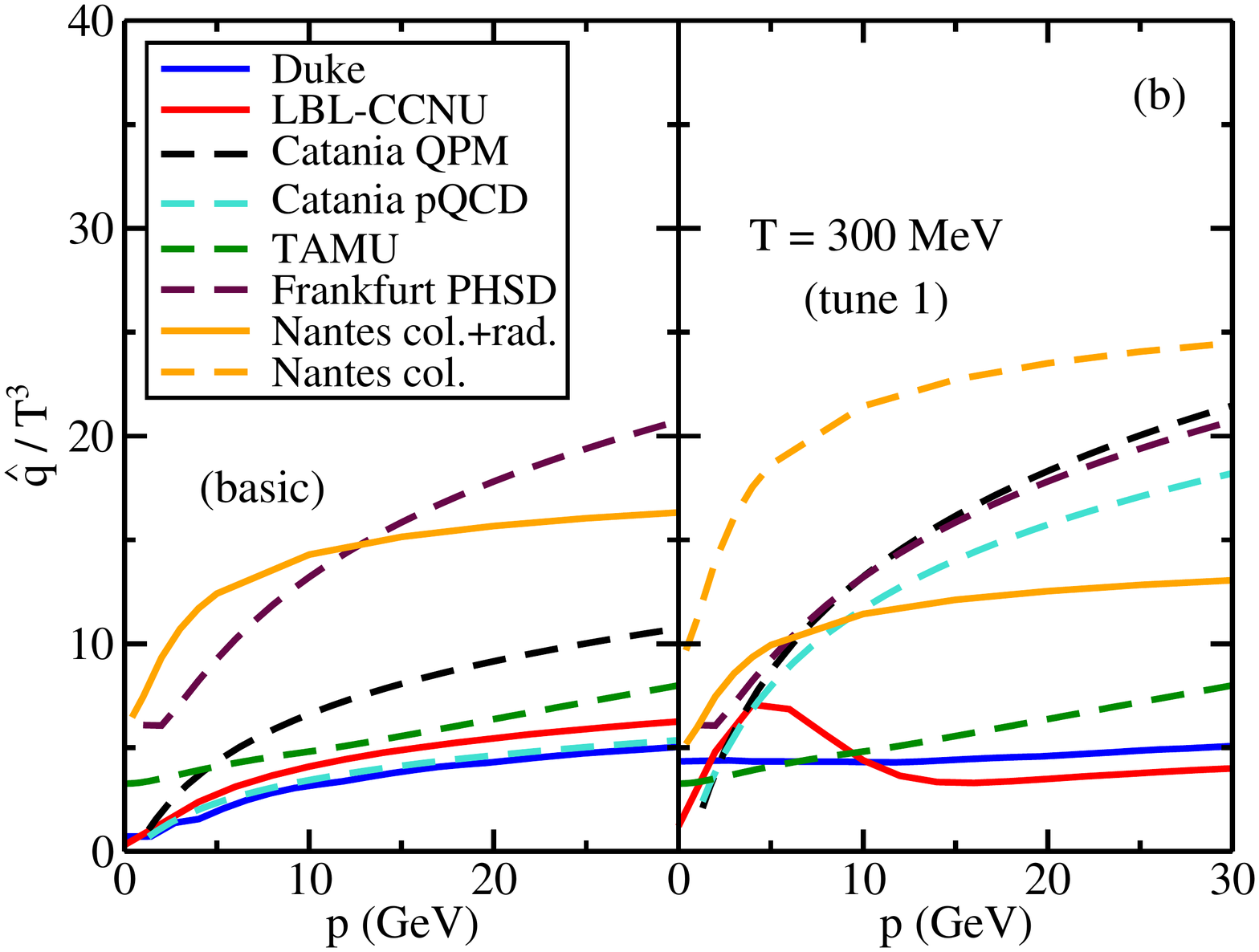}\label{fig:TC-p-qhat}}
        \caption{The momentum dependence of (a) $A$ and (b) $\hat{q}/T^3$. Left columns (``basic") are direct calculations from different models; and right (``tune 1") are extracted from comparing to data with different models.}
        \label{fig:TC-p}
\end{figure}

\begin{figure}
        \centering
             \subfigure{\includegraphics[width=0.45\textwidth]{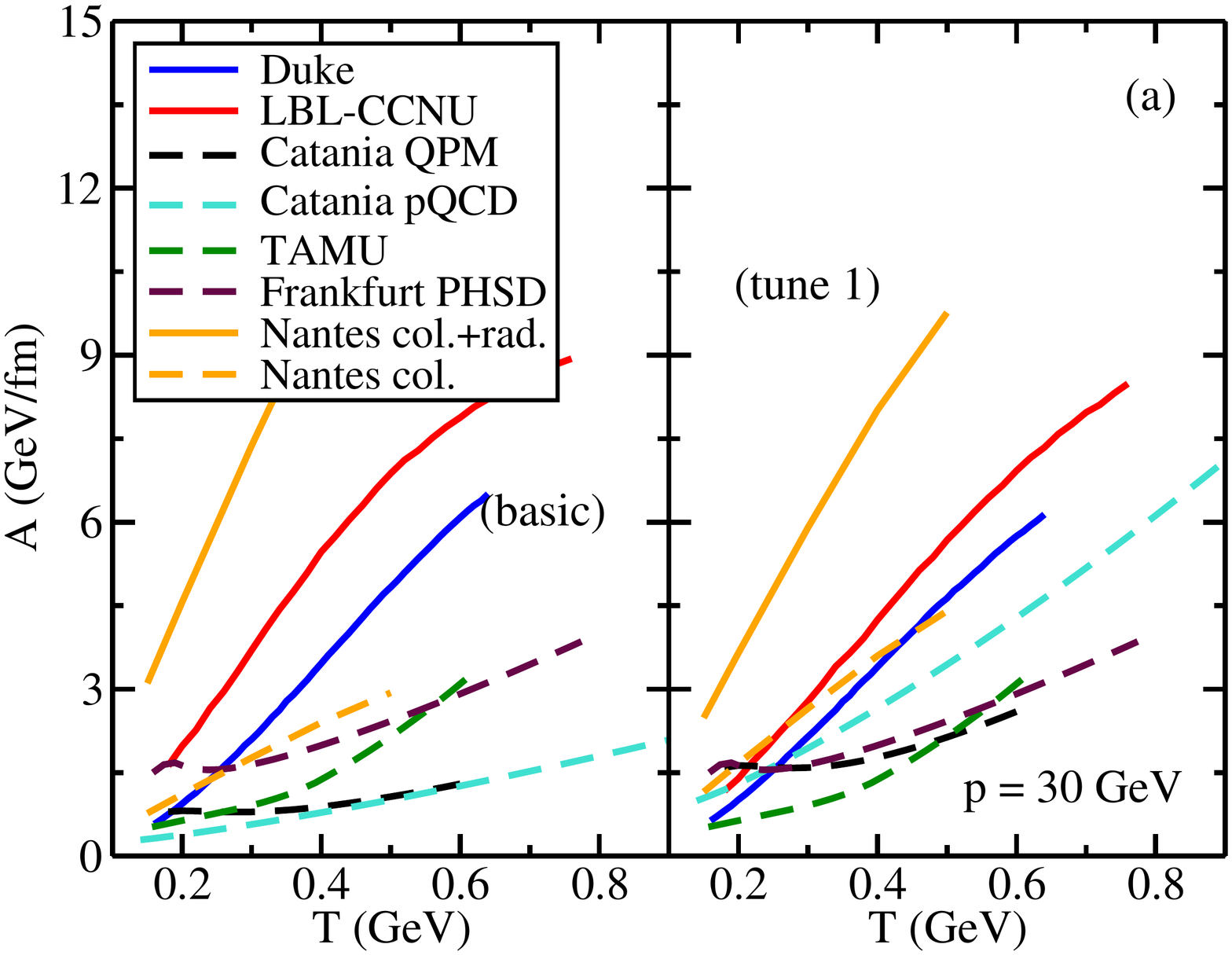}\label{fig:TC-T-drag}}\vspace{-20pt}
             \subfigure{\includegraphics[width=0.45\textwidth]{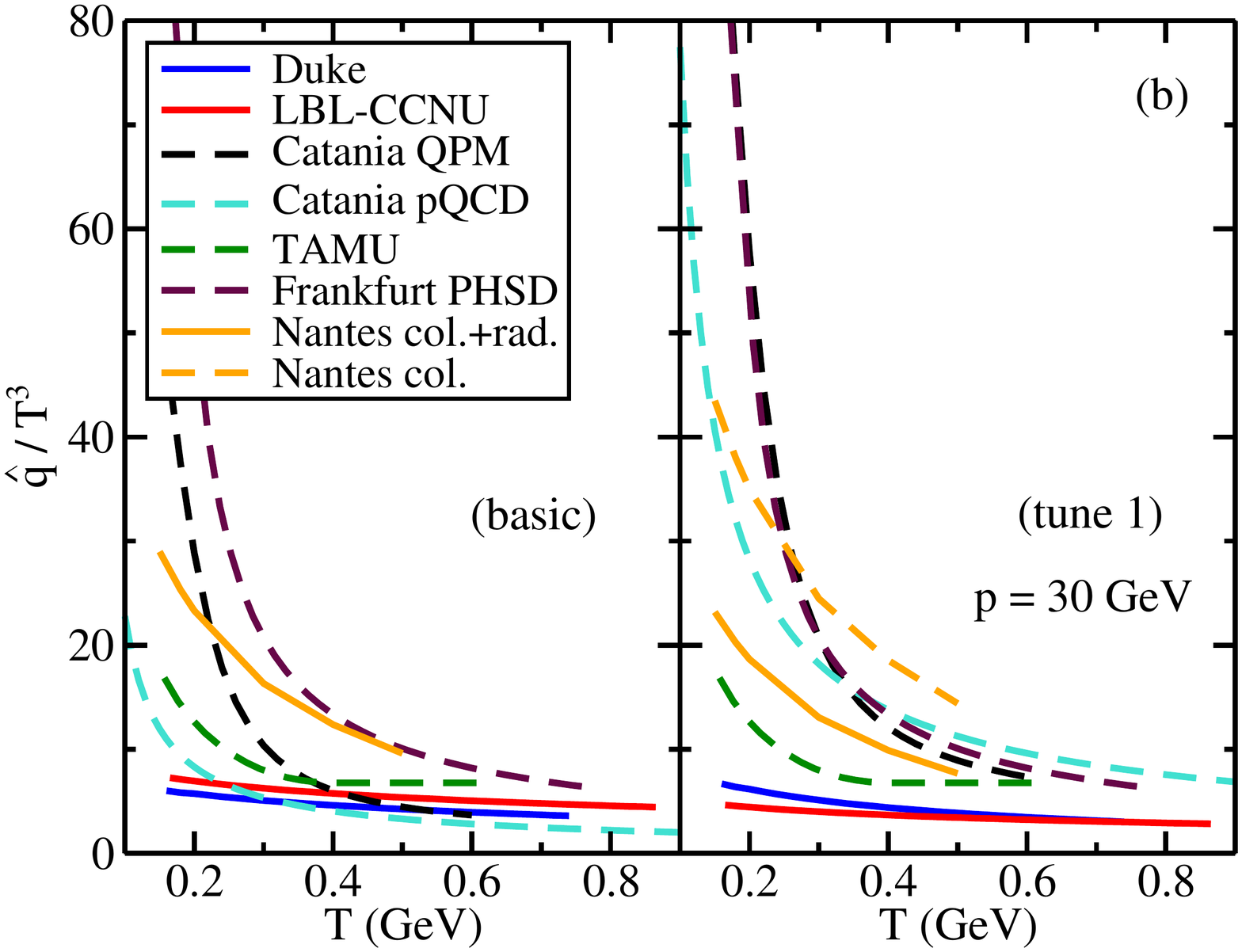}\label{fig:TC-T-qhat}}
        \caption{The temperature dependence of (a) $A$ and (b) $\hat{q}/T^3$. Left columns (``basic") are direct calculations from different models; and right (``tune 1") are extracted from comparing to data with different models.}
        \label{fig:TC-T}
\end{figure}

\begin{table*}[t]
\begin{tabular}{l | c | c | c | c | c}
Models & \;transport schemes\; & basic & tune 1 & tune 2 \\
\hline \hline 
Duke & Langevin & \;fixed $\alpha_\mathrm{s}=0.3$ only\; & \;$(\alpha, \beta, \gamma)=(1.89, 1.59, 0.26)$\; & \;$D_\mathrm{s}$ = $0.77\times D_\mathrm{s}$(tune 1)\; \\ 
\hline
CCNU-LBNL & Boltzmann & \;fixed $\alpha_\mathrm{s}=0.3$ only\; & $\alpha_s=0.24$ with $K_p$ & $\alpha_s=0.28$ with $K_p$  \\ 
\hline
Catania QPM & Boltzmann & running $\alpha_s (T)$ & $K=2.25$ & $K=3.45$  \\
\hline
Catania pQCD & Langevin & running $\alpha_s (T)$ & $K=3.4$ &  $K=3.1$ \\
\hline
TAMU & Langevin & $U$ from lQCD  & no tuning & $K=2.45$   \\
\hline
Frankfurt PHSD & Boltzmann & running $\alpha_s (T)$  & no tuning &  $K=1.6$ \\
\hline
Nantes col. + rad. & Boltzmann & running $\alpha_s (q^2)$ & $K=0.8$ & $K=0.45$ \\
\hline
Nantes col. only & Boltzmann & running $\alpha_s (q^2)$ & $K=1.5$ & $K=1.1$
\end{tabular}
\caption{Key inputs and model tunings of different HQ transport formalisms. In CCNU-LBNL model tune 1 and tune 2, a momentum-dependent $K_p$ factor is applied in addition to the fixed coupling constant $\alpha_\mathrm{s}$ as discussed in Sec.~\ref{subsec:Berkeley}. }
\label{tab:tuning}
\end{table*}

With model parameters adjusted in order to describe the experimental data on the nuclear modification factor for $D$ mesons, one can evaluate the HQ transport coefficients in each model. In Fig.~\ref{fig:TC-p} and \ref{fig:TC-T}, we compare the transport coefficients between different model approaches. Solid lines are used for models that include both elastic and inelastic processes in this study, while dashed lines are for models that only include elastic scatterings. In Fig.~\ref{fig:TC-p}, we compare the drag $A$ and transport coefficient $\hat{q}$ as functions of HQ momentum in a thermal medium with a fixed temperature of $T=300$~MeV; and in Fig.~\ref{fig:TC-T}, we compare them as functions of the medium temperature with a fixed HQ momentum of $p=30$~GeV/$c$. In each figure, the left column corresponds to transport coefficients directly calculated from different models without tuning (denoted as ``basic"), while the right columns represent the extracted transport coefficients after the model calculations are calibrated to the experimental data of the $D$ meson nuclear modification factor $R_\mathrm{AA}$ within the $p_\mathrm{T}$ range of $2\sim 15$~GeV/$c$ in central Pb-Pb collisions at 2.76~ATeV in Fig.~\ref{fig:data-LHC} (denoted as ``tune 1"). The key inputs of the different models and their parameter tunings are summarized in Table~\ref{tab:tuning}. One may refer to Sec.~\ref{sec:models} and references therein for more detailed descriptions of each model. 

As shown in Table~\ref{tab:tuning} (column labeled ``basic"), different assumptions about HQ-medium interactions are adopted in different model setups: Duke and CCNU-LBNL assume a fixed coupling constant $\alpha_\mathrm{s}$ in calculating transport coefficients; Catania and Frankfurt (PHSD) assume a temperature dependent $\alpha_\mathrm{s}$; Nantes assumes a momentum-transfer dependent $\alpha_\mathrm{s}$; and TAMU utilizes the internal energy extracted from lQCD to describe heavy-light quark interactions. This leads to a clear separation of the directly calculated HQ transport coefficients as shown in the left (``basic") columns of Figs.~\ref{fig:TC-p} and~\ref{fig:TC-T}. 

To more quantitatively describe the $D$-meson nuclear modification factor $R_\mathrm{AA}$, certain parameters in the model calculations need to be adjusted. As shown in  Tab.~\ref{tab:tuning} (column of ``tune 1"), Duke introduces ($\alpha$, $\beta$, $\gamma$) to parametrize the non-perturbative part of the diffusion coefficient (see Sec.~\ref{subsec:Duke}) which are then calibrated using a Bayesian method; CCNU-LBNL needs to adjust the coupling constant together with a momentum dependent $K$-factor (see Sec.~\ref{subsec:Berkeley}) that models the non-perturbative contribution to $\hat{q}$; Catania and Nantes apply a constant $K$ factor on the overall HQ scattering cross section to include physics beyond the current model descriptions, cf.~Secs.~\ref{subsec:Catania} and~\ref{subsec:Nantes}. The TAMU (Sec.~\ref{subsec:TAMU}) and Frankfurt (Sec.~\ref{subsec:Frankfurt}) models present direct calculations without tuning when comparing to experimental data. 

If one assumes that transport coefficients can effectively quantify the HQ energy loss inside the QGP, one would expect convergence of the extracted drag $A$ and jet transport parameter $\hat{q}$ once different models are simultaneously calibrated to the experimental data on $R_\mathrm{AA}$. However, this is not the case as indicated by the right columns (``tune 1")  of Figs.~\ref{fig:TC-p} and~\ref{fig:TC-T}.  At high momenta, the results spread over more than a factor of 5 even if the different models provide comparable values of the $D$-meson $R_\mathrm{AA}$. This apparently calls for a systematical comparison between various model calculations in order to understand the different mechanisms that affect the $R_{\rm AA}$ and narrow down the uncertainties of the extracted transport coefficients.

\subsection{Nuclear modification of charm quarks in a brick}
\label{subsec:backToBrick}

\begin{figure}
        \centering
                \includegraphics[width=0.45\textwidth]{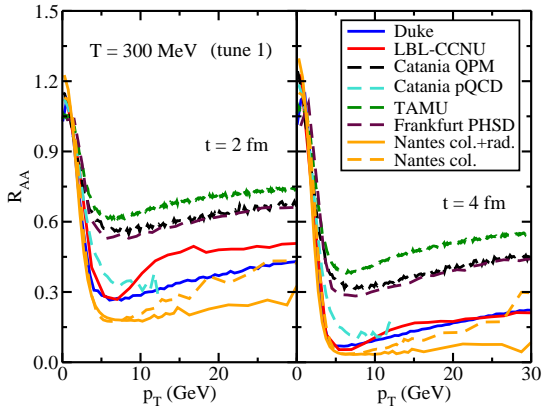}
        \caption{The nuclear modification factor of charm quark in a static medium at $t=2$~fm and 4~fm, using the ``tune 1" setup in each model that provides the best description of experimental data.}
        \label{fig:plot-RAA-tune1-T300}
\end{figure}

The wide variation of the extracted transport coefficients as presented in the previous section, different ingredients may contribute.  They include the initial spectra of hard scatterings, formation times of the heavy quarks,  the treatment of the bulk (QGP) medium, the formalism for hadronization converting heavy quarks into HF hadrons, etc.. The initial HQ spectra are usually constrained by the measured $D$ meson spectra in p-p collisions. But for the other processes, one has to take a more systematic approach. To eliminate the above differences as possible sources for the divergent transport coefficients extracted from the model calculations, and to search for a direct correlation between transport coefficients and HQ energy loss, we design in the following a so-called "QGP brick" calculation. First, we initialize charm quarks with a simplified power-law parametrization of the $p_\mathrm{T}$ sectra that is inspired by perturbative calculations~\cite{Moore:2004tg},
\begin{equation}
\label{eq:spectra}
\frac{dN}{d^2 p_\mathrm{T}}\propto \frac{1}{(p^2_\mathrm{T}+\Lambda^2)^\alpha}
\end{equation}
with $\alpha=3.52$ and $\Lambda=1.85$~GeV. Then we let charm quarks evolve through a brick medium at a fixed temperature for a given time of propagation. The final-state spectra are analyzed at the partonic level at the end of the evolution to exclude uncertainties from different hadronization schemes.

With this setup, we  calculate suppression factor $R_\mathrm{AA}$ with each model (``tune 1") (transport coefficients extracted from different models are  shown as  ``tune 1" in Figs.~\ref{fig:TC-p} and~\ref{fig:TC-T}) for a charm quark traveling through the brick at a constant temperature of $T=300$~MeV. The results are shown in Fig.~\ref{fig:plot-RAA-tune1-T300}. One observes an apparent difference in $R_\mathrm{AA}$ at the time $t=2$~fm and at 4~fm of the charm-quark propagation, although these models are tuned (``tune 1") to reproduce the experimental data on $R_\mathrm{AA}$ for $D$ meson in central Pb+Pb collisions at the LHC in the calculations of their original frameworks. This implies that there must be significant differences in the bulk evolution and hadronization in these models, and that their effects on the final charm-meson spectra lead to large variations of the extracted drag $A$ and jet transport coefficient $\hat{q}$, even though they are tuned to fit the experimental data in heavy-ion collisions.

By comparing different model results in Fig.~\ref{fig:plot-RAA-tune1-T300} and the ``tune 1" column of Figs.~\ref{fig:TC-p-drag} and~\ref{fig:TC-T-drag}, one may observe that the general correlations between the drag coefficient and charm-quark suppression due to energy loss still remain as expected: As $A$ increases, so does the energy loss the heavy quarks suffer, and thus the value of $R_\mathrm{AA}$ becomes smaller. For instance, the lowest value of $A$ in Figs.~\ref{fig:TC-p} and~\ref{fig:TC-T-drag} (tune 1) is obtained by TAMU whereas the highest values are obtained by Nantes (with gluon radiation), which translates into the largest and smallest $R_\mathrm{AA}$ values in Fig.~\ref{fig:plot-RAA-tune1-T300}, respectively. The other results lie in between. Note that the ordering here only reflects general features of the correlation between $A$ and $R_\mathrm{AA}$. Other details in each model, such as the energy loss fluctuations, may affect this  hierarchy as well.

\begin{figure}
        \centering
                \includegraphics[width=0.45\textwidth]{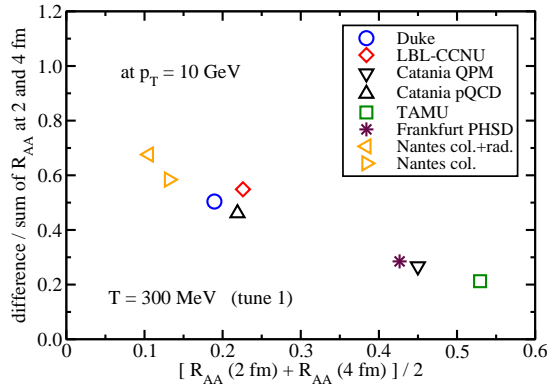}
        \caption{Correlation between the heavy quark $R_\mathrm{AA}$ and the ``anisotropy" extracted from the $R_\mathrm{AA}$ at 2~fm and 4~fm in a static medium.}
        \label{fig:plot-v2-from-RAA}
\end{figure}

Although the elliptic flow coefficient $v_2$ cannot be directly defined within a brick medium, we are still able to investigate a proxy in terms of the asymmetry of charm-quark energy loss through different path length as shown in Fig.~\ref{fig:plot-v2-from-RAA}, where we plot the ratio between the difference and the sum of the charm-quark $R_\mathrm{AA}$ at $t=2$~fm/$c$ and $t=4$/$c$~fm, as a function of their average value. The $R_\mathrm{AA}$ is evaluated at $p_\mathrm{T}=10$~GeV from Fig.~\ref{fig:plot-RAA-tune1-T300} in each curve. The $y$-axis of Fig.~\ref{fig:plot-v2-from-RAA} mimics the value of $v_2$ due to the asymmetric energy loss through different path lengths (2~fm vs 4~fm), and the $x$-axis quantifies the average energy loss. Although the six model calculations give different values of charm-quark $R_\mathrm{AA}$ in a brick, Fig.~\ref{fig:plot-v2-from-RAA} displays a clear correlation between the average energy loss and the energy loss asymmetry due to different path lengths, consistent with the expectation that $v_2$ decreases as $R_\mathrm{AA}$ increases. Note that apart from the energy loss asymmetry due to different path lengths, The HQ $v_2$ in heavy-ion collisions is also influenced by the collective flow of the expanding medium. This effect is not considered in the discussion here.

\section{Narrowing down the uncertainty of the extracted transport coefficients utilizing a brick}
\label{sec:brickProblem}

\subsection{Common baseline within a brick}
\label{subsec:brickSetup}

\begin{figure}
        \centering
                \includegraphics[width=0.45\textwidth]{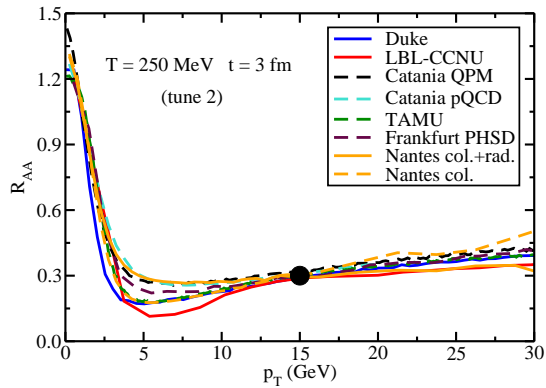}
        \caption{Common baseline of charm quark $R_\mathrm{AA}$ in a brick with model parameters denoted as ``tune 2" in this work.}
        \label{fig:plot-RAA_tune2_t3}
\end{figure}

In the previous section, we showed that the $R_\mathrm{AA}$ for heavy quarks is quite different for the different models. This indicates  that different evolutions of the bulk medium and different hadronization schemes are at the origin of the theoretical uncertainties in extracting the HQ transport coefficients from existing transport models. To eliminate these model uncertainties, we design a common baseline within our simple brick setup, from which we extract and compare the transport coefficients between the different transport models. 

In this common baseline, we first readjust parameters in each model such that charm quarks with an initial spectrum as given in Eq.~(\ref{eq:spectra}) 
have a suppression factor $R_\mathrm{AA}=0.3$ at $p_\mathrm{T} =15$~GeV/$c$ after they propagate through a static brick at a constant temperature $T=250$~MeV for $t=3$~fm/$c$. The $R_{\mathrm{AA}}$ values as a function of $p_\mathrm{T}$ at t=3 fm/$c$ and $T =250$~ MeV  are shown in Fig.~\ref{fig:plot-RAA_tune2_t3} and agree quite reasonably for the different transport approaches, especially at large $p_\mathrm{T}$. 
This common baseline suggests that $T=250$~MeV should be a reasonable approximation of the average temperature over an average distance of 3 fm in the QGP in realistic Pb+Pb collisions, and $R_\mathrm{AA}=0.3$ is approximately the experimental value on $D$-meson suppression in central Pb-Pb collisions [recall Fig.~\ref{fig:data-LHC}] around $p_\mathrm{T} =15$~GeV/$c$ where the difference of $R_{\mathrm{AA }}$ between charm quark and $D$ meson should be small.

The model parameters tuned to this common baseline are summarized as ``tune 2" in Tab.~\ref{tab:tuning}. Compared to ``tune 1", where the original full models for realistic heavy-ion collisions are tuned to fit the experimental data on charmed meson suppression, the Duke, CCNU-LBNL, Catania-QPM, TAMU and Frankfurt models need to increase the HQ-medium interaction by either decreasing the spatial diffusion coefficient, increasing the coupling constant $\alpha_\mathrm{s}$, or applying a $K>1$ factor to the overall scattering cross section. On the other hand, Catania-pQCD and Nantes models need to decrease the interaction by using smaller $K$ factors. This also suggests the underlying differences in the transport implementation (Langevin vs. Boltzmann) and bulk evolution adopted by different groups.

\begin{figure}
        \centering
                \includegraphics[width=0.45\textwidth]{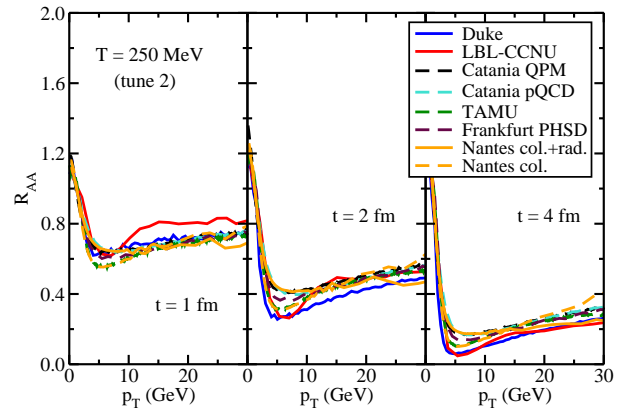}
        \caption{Time evolution of charm quark $R_\mathrm{AA}$ within the common brick.}
        \label{fig:plot-RAA_tune2_t246}
\end{figure}

With the ``tune 2" parameters fixed by the charm-quark $R_\mathrm{AA}$ in a static brick at $t=3$~fm/$c$, the agreement of the $R_\mathrm{AA}$ at other times, $t=1$, 2 and 4 fm/$c$, is also reasonable, see Fig.~\ref{fig:plot-RAA_tune2_t246} . However, a closer examination reveals more detailed insights into different features of the  models. For instance, the inelastic energy loss implemented in Duke and CCNU-LBNL approaches is based on the higher-twist energy loss formalism, in which the medium-induced gluon radiation rate increases with time due to the Landau-Pomeranchuk-Migdal (LPM) interference between the soft HQ-medium scattering and the initial hard scattering of HQ production. This time dependence of HQ energy loss is not included in other models here. Therefore, with $R_\mathrm{AA}$=0.3 fixed at $t=3$~fm/$c$, the $R_\mathrm{AA}$ values from Duke and CCNU-LBNL are slightly larger than other models at earlier time but slightly smaller at later time. Note that after including such time dependent inelastic processes, the drag coefficient $A$ also increases with time. In order to compare with other model calculations, $A$ from Duke and CCNU-LBNL in this work represents the average value within the first 3~fm/$c$.

\subsection{Consistency of the extracted transport coefficients}
\label{subsec:brickResults}

\begin{figure}
        \centering
             \subfigure{\includegraphics[width=0.45\textwidth]{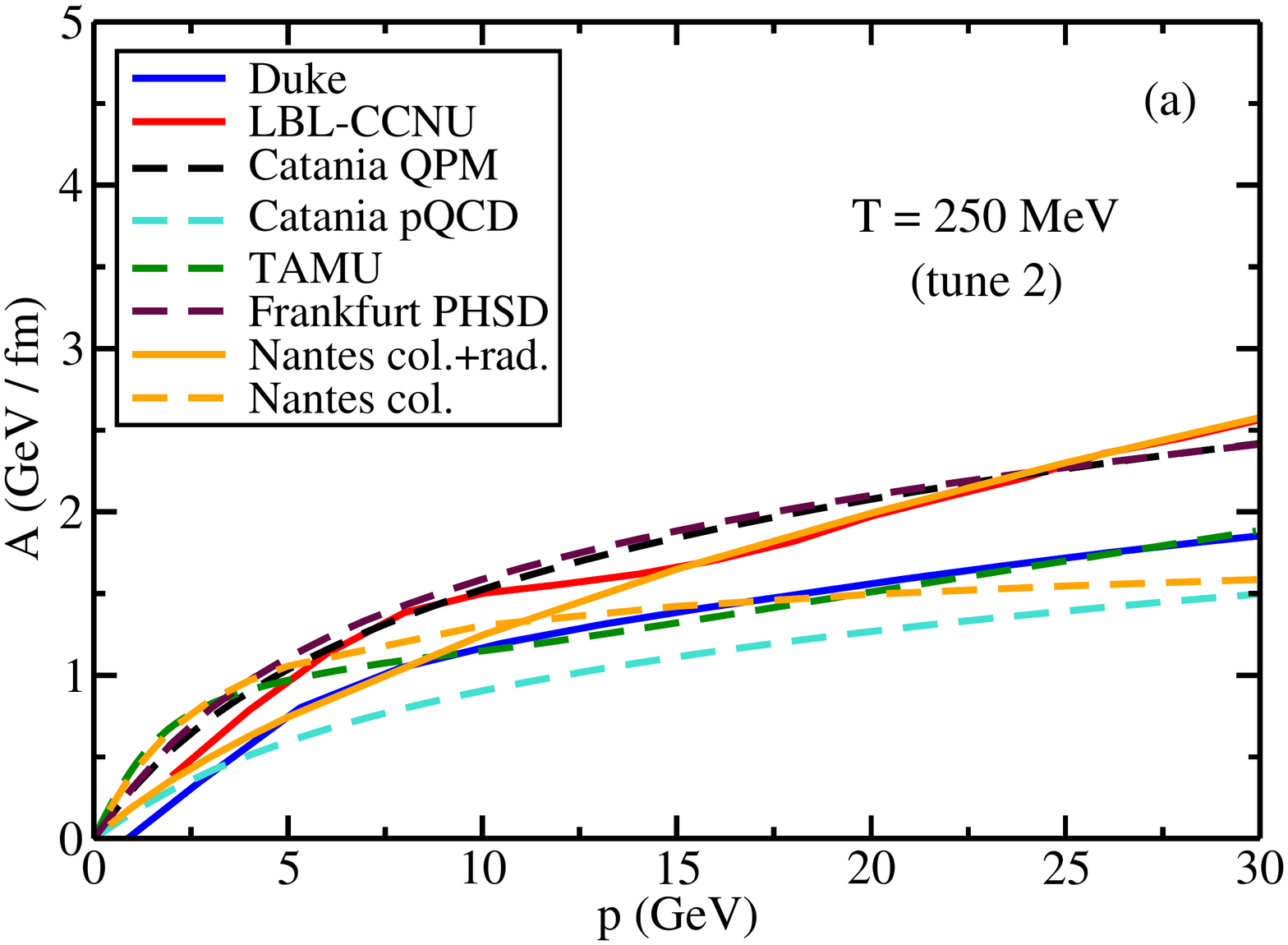}\label{fig:plot-tune2-dragTot-p}}\vspace{-20pt}
             \subfigure{\includegraphics[width=0.45\textwidth]{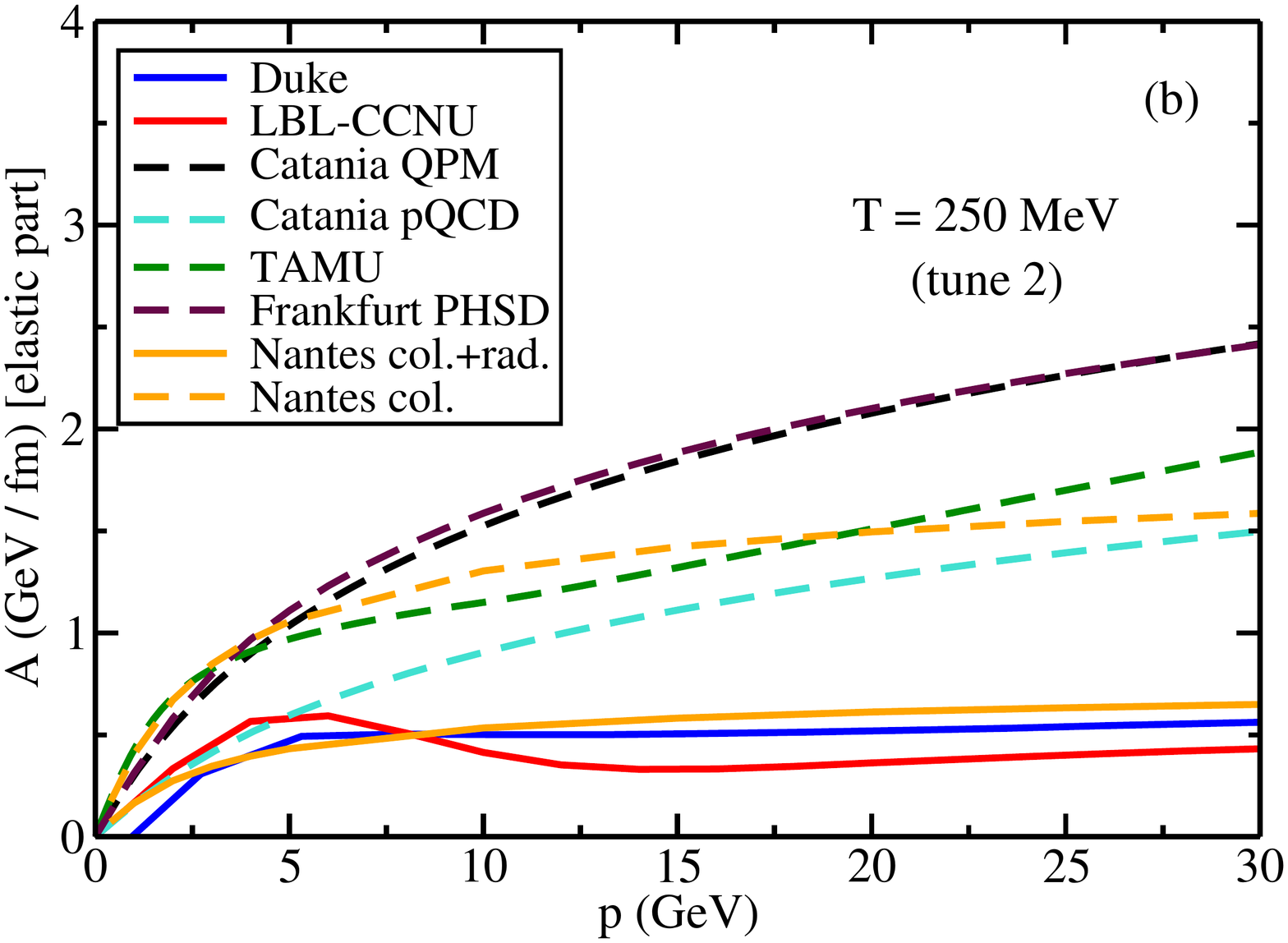}\label{fig:plot-tune2-drag-p}}\vspace{-20pt}
             \subfigure{\includegraphics[width=0.45\textwidth]{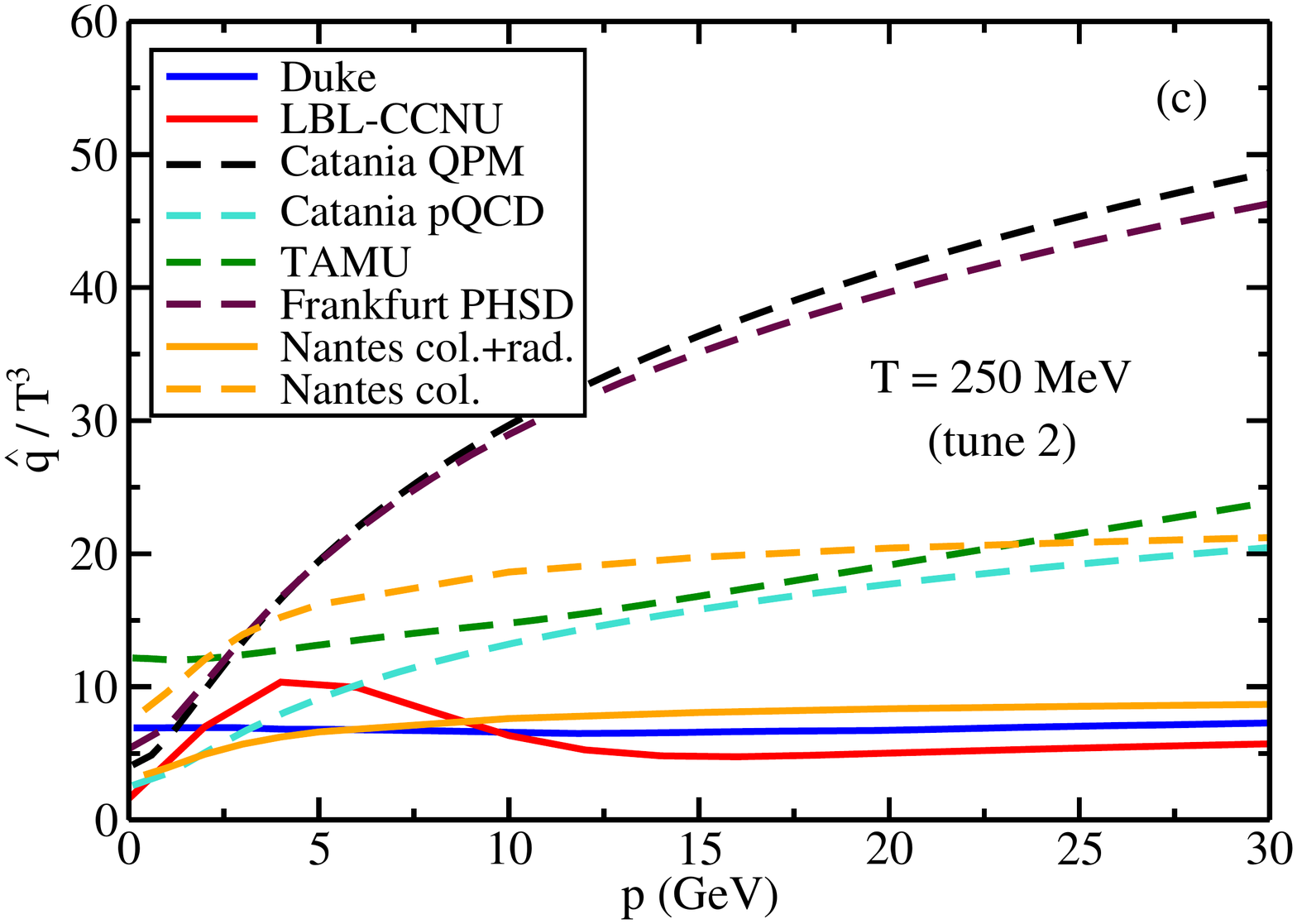}\label{fig:plot-tune2-qhat-p}}
        \caption{Transport coefficients extracted from the common baseline of $R_{\rm AA}$ in a brick:  (a) $A$, (b) elastic contribution to $A$ and (c) $\hat{q}/T^3$.}
        \label{fig:TC-tune2-T250}
\end{figure}

After calibrating the various model calculations to our common baseline within a brick, we present the extracted transport coefficients in a QGP at $T=250$~MeV as functions of the charm-quark momentum in Fig.~\ref{fig:TC-tune2-T250}. Figure~\ref{fig:plot-tune2-dragTot-p} displays the total drag coefficient $A$ (elastic+inelastic for the models that include both processes), Fig.~\ref{fig:plot-tune2-drag-p} displays the elastic contribution to $A$, and Fig.~\ref{fig:plot-tune2-qhat-p} displays the elastic transport parameter $\hat{q}$. With the same conventions as in Sec.~\ref{sec:coefficient}, solid lines correspond to calculations that include both elastic and inelastic processes, while for the dashed lines only elastic scattering is included.

\begin{figure}
        \centering
            \subfigure{\includegraphics[width=0.45\textwidth]{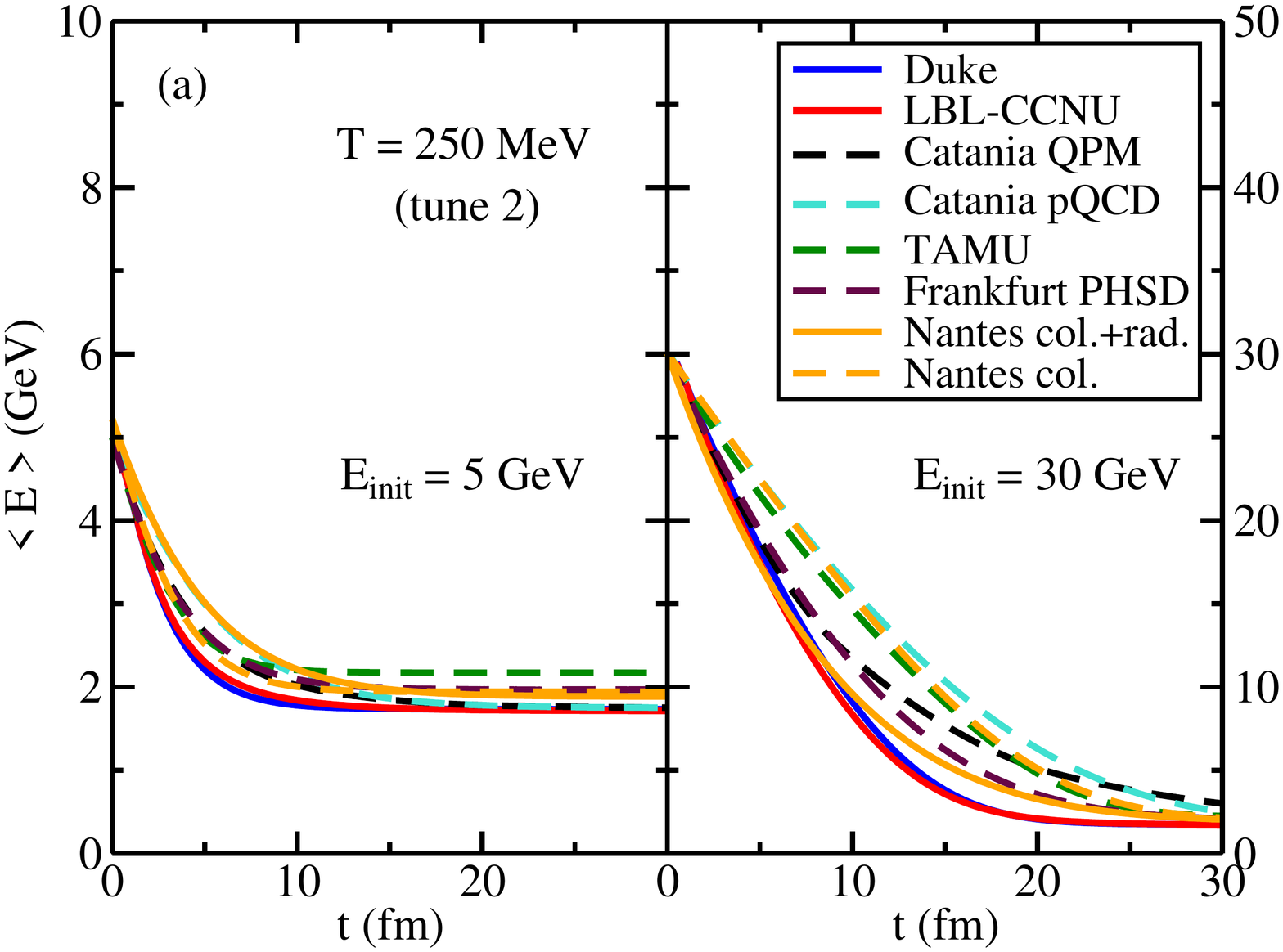}\label{fig:plot-tune2-eAvr}}\vspace{-20pt}
            \subfigure{\includegraphics[width=0.45\textwidth]{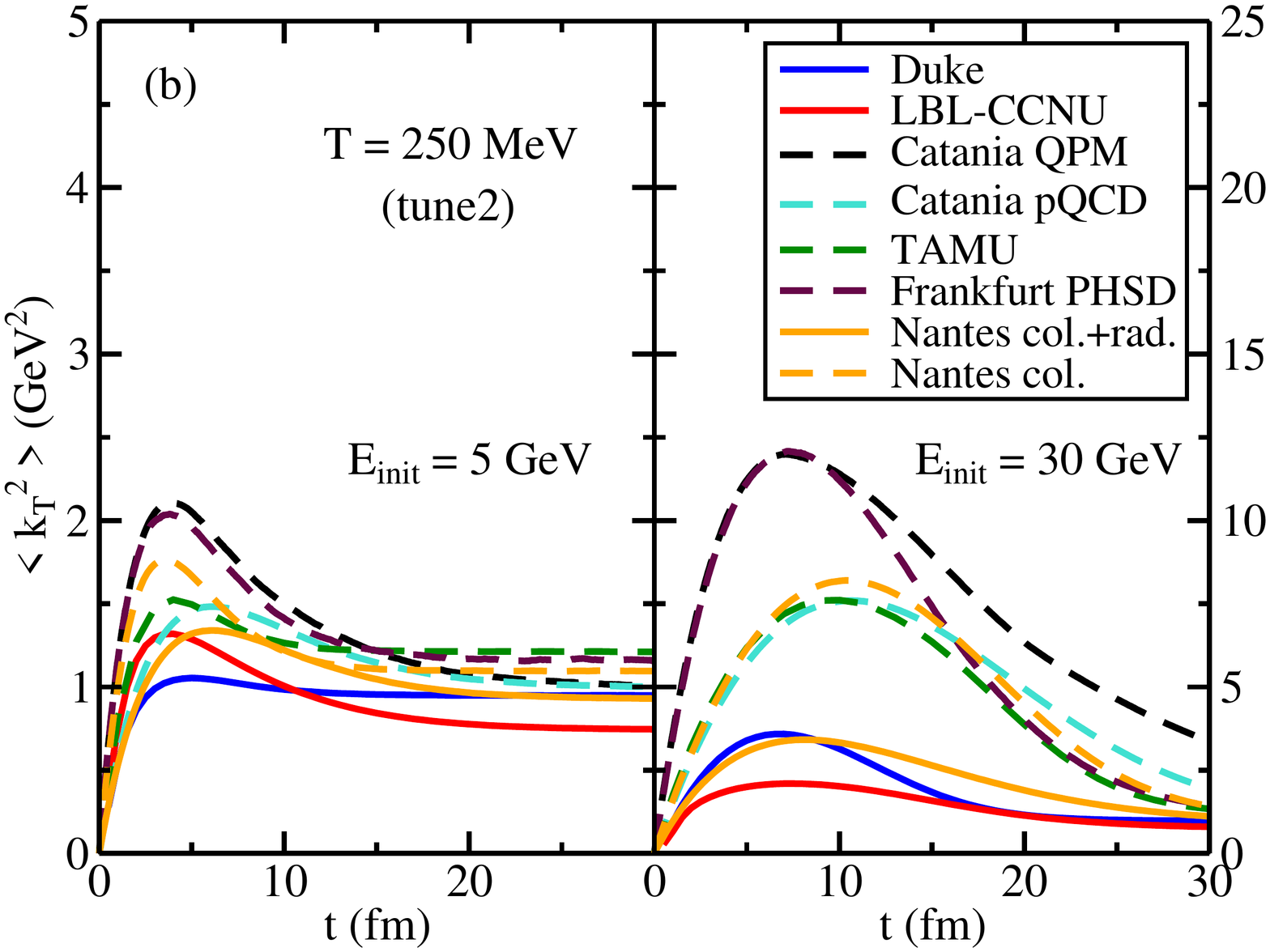}\label{fig:plot-tune2-pT2Avr}}\vspace{-20pt}
            \subfigure{\includegraphics[width=0.45\textwidth]{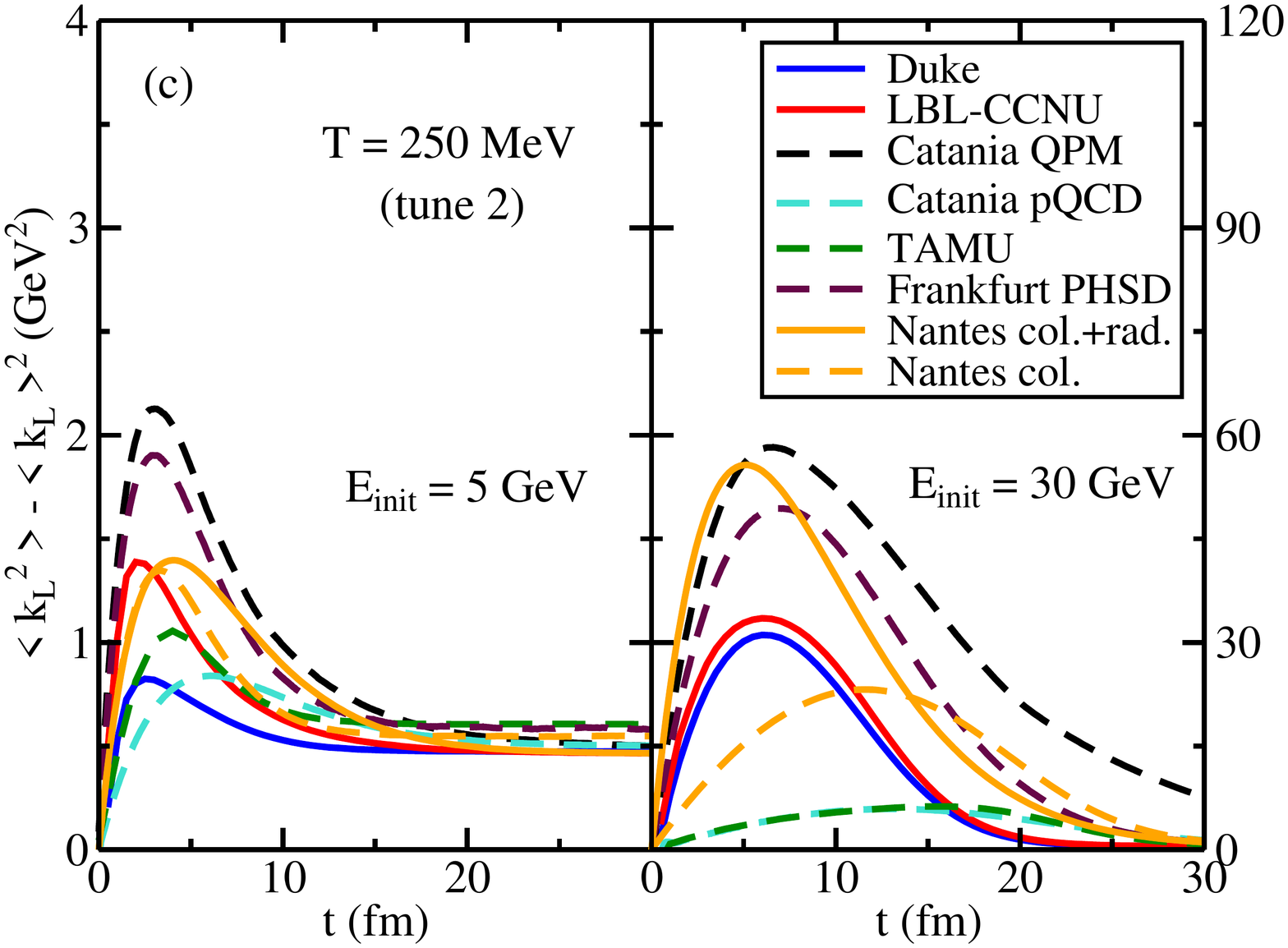}\label{fig:plot-tune2-dpL2Avr}}
        \caption{Time evolution of (a) average energy, (b) average transverse momentum broadening and (c) longitudinal momentum fluctuation of heavy quarks inside a common static medium.}
        \label{fig:tune2-average}
\end{figure}

With this brick setup, one observes in Fig.~\ref{fig:plot-tune2-dragTot-p} that the drag coefficients extracted from different models become similar, within a factor of about 2 in variation, significantly smaller as compared to that in Figs.~\ref{fig:TC-p} and \ref{fig:TC-T}. The remaining differences between the results of the different models come from the fact that the HQ $R_\mathrm{AA}$ is determined not only by the average energy loss, or drag, but also by the fluctuation of the energy loss as well as interference effects. For instance, the spectrum of medium-induced gluon radiation in inelastic processes is different from the distribution of energy transfer in elastic scatterings, and therefore may lead to the separation of the extracted $A$ between models with pure elastic energy loss and models that include gluon emission. Within elastic scatterings, the relation between drag and diffusion can also vary due to the different treatment of the thermal scattering partners. The QPM and PHSD models, which use rather large thermal parton masses when approaching $T_c$, lead to larger transverse and especially longitudinal fluctuations (as will be shown in Fig.~\ref{fig:tune2-average}). The ensuing larger fluctuations then require a larger drag to accommodate a given $R_\mathrm{AA}$.

In Figs.~\ref{fig:plot-tune2-drag-p} and \ref{fig:plot-tune2-qhat-p}, we find that the elastic part of the transport coefficients from different models fall into three groups: (1) approaches that incorporate both elastic and inelastic energy loss (Duke, CCNU-LBNL and Nantes col.+rad.); (2) approaches that contain pQCD-driven energy loss via elastic scattering off partons with small quasi-particle masses (Catania-pQCD, TAMU\footnote{The TAMU approach is non-perturbative, but is consistent with perturbative results at large momentum scales.} and Nantes-coll.-only); and (3) approaches based on elastic scatterings driven by quasi-particle models with large masses especially near the phase transition region (Catania-QPM and Frankfurt-PHSD).

To better illustrate the differences between various models, we present the time evolution of the average energy, transverse momentum squared and longitudinal momentum fluctuations of heavy quarks in Fig.~\ref{fig:tune2-average}. Here, charm quarks are initialized with a fixed momentum (5~GeV/$c$ for the left columns and 30~GeV/$c$ for the right columns), and then evolved through a static medium with $T=250$~MeV.  Although separations between different approaches are small at low momenta (5~GeV/$c$), which is more relevant to the determination of the diffusion coefficient $D_s$, differences for high energy (30~GeV/$c$) charm quarks, more relevant to the determination of $\hat{q}$, are evident. As expected, one can observe in Fig.~\ref{fig:plot-tune2-eAvr} that approaches within group (1) and (3) give a faster HQ energy loss and approach to thermalization than group (2) due to the larger drag coefficients in the former. In Fig.~\ref{fig:plot-tune2-pT2Avr}, we furthermore find that the three groups of approaches result in a different amount of transverse-momentum broadening. Charm quarks within group (1) accumulate the least amount of $p_\mathrm{T}$ broadening, since the medium-induced gluon emission prefers collinear emission with respect to the parent heavy quark and thus is less effective in changing its direction compared to elastic scattering. Between the two groups of pure elastic approaches, group (3) (heavy quasi-particles) generates larger transverse momentum broadening than group (2) (pQCD-based models). With the similar amount of longitudinal momentum loss, the transverse-momentum transfer in group-3 models is larger because of both the heavier thermal masses and the larger Debye screening masses employed there. 
These findings motivate the study of the angular correlations between HF pairs for the future, providing more constraints on the properties of the HQ energy loss mechanism~\cite{Nahrgang:2013saa,Cao:2015cba}. The variation in the longitudinal momentum fluctuations as shown in Fig.~\ref{fig:plot-tune2-dpL2Avr} should also lead to variations of the final HQ suppression and the extraction of the pertinent transport coefficients.

\subsection{Non-trivial temperature dependence of transport coefficients}
\label{subsec:TDependence}

\begin{figure}
        \centering
            \subfigure{\includegraphics[width=0.45\textwidth]{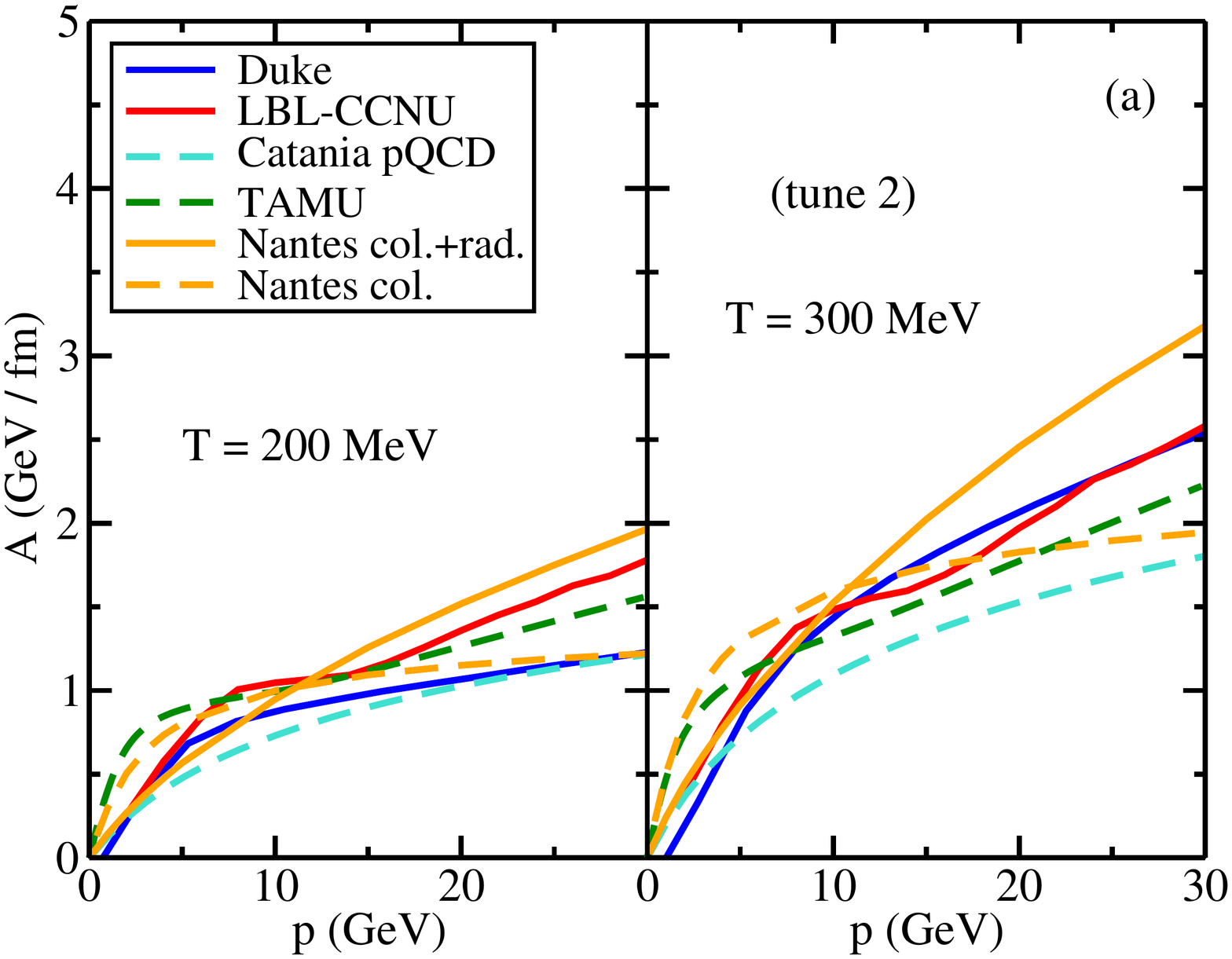}\label{fig:plot-tune2-dragTot-p-otherT}}\vspace{-20pt}
            \subfigure{\includegraphics[width=0.45\textwidth]{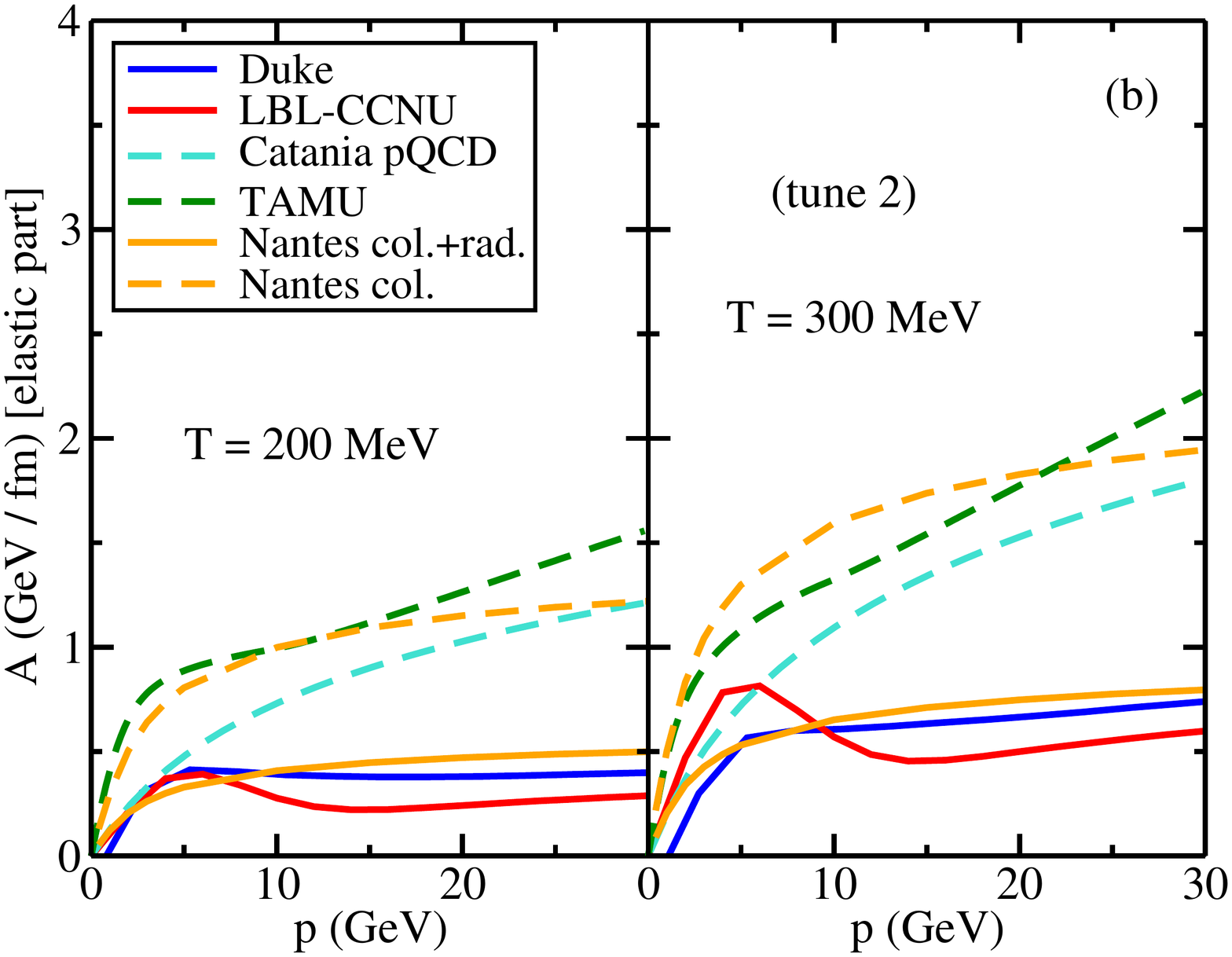}\label{fig:plot-tune2-drag-p-otherT}}\vspace{-20pt}
            \subfigure{\includegraphics[width=0.45\textwidth]{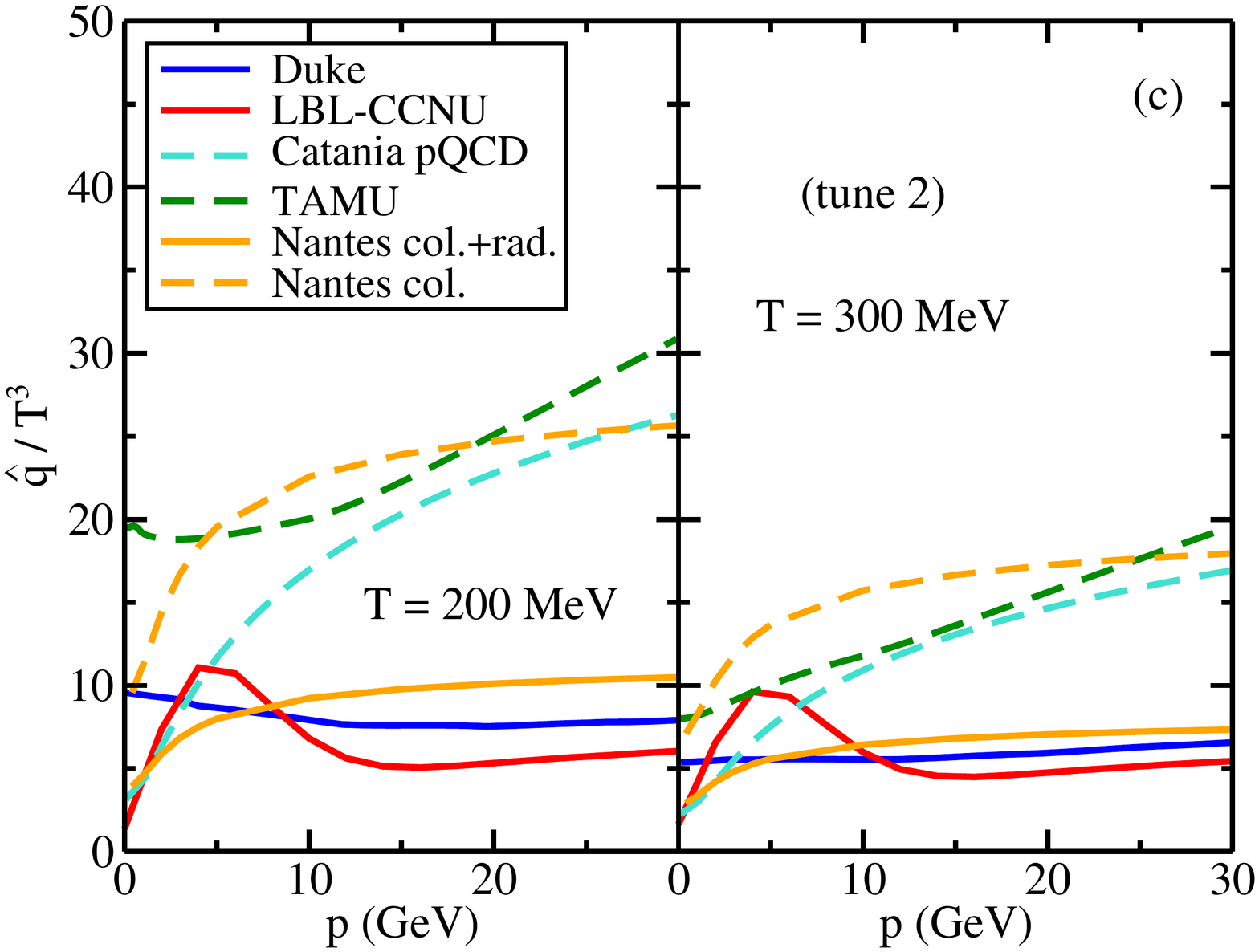}\label{fig:plot-tune2-qhat-p-otherT}}
        \caption{Calculations of (a) $A$, (b) elastic contribution to $A$ and (c) $\hat{q}/T^3$, at $T=200$~MeV and 300~MeV with ``tune 2".}
        \label{fig:TC-tune2-otherT}
\end{figure}

Finally, in Fig.~\ref{fig:TC-tune2-otherT},  we show charm quark transport coefficients from different models with ``tune 2" at different temperatures. For a clearer presentation, we only include $A$ and $\hat{q}$ from pQCD driven models, from  group (1) -- elastic + inelastic scattering, and group (2) -- elastic scattering alone. As one can see, convergence between different approaches within each group still holds up at $T=200$~MeV and 300~MeV. However, compared to the results in Fig.~\ref{fig:TC-tune2-T250}, we also observe an increasing divergence within each group when we deviate from $T=250$~MeV, the temperature where our common baseline is defined. This results from the different temperature dependences of HQ transport coefficients in the transport models that are used in this study; its origins include different temperature dependences of the light-quark masses, coupling constants, and Debye masses.

\subsection{Future improvements of heavy quark transport models}
\label{sec:future}

With the comparisons presented in this work, different groups have assessed their current model performance and prepared to further improve these models in several directions. 

Currently the Catania Boltzmann approach is set up only considering energy loss by elastic collisions of heavy quarks with bulk partons, because it mainly focuses on heavy quark dynamics at low momentum. In this work, it is found that for the common reference point ($R_\mathrm{AA} =0.3$ at $p_\mathrm{T}\sim 15$~GeV) we design, radiative energy loss could play a significant role. This radiative process will be included in the Catania model. As a preliminary attempt, the Catania Boltzmann approach has adopted the higher-twist formalism for medium-induced gluon radiation that is currently implemented in the Duke model and CCNU-LBNL model, and verified the sizable contribution of radiative energy loss on heavy quark $R_\mathrm{AA}$ and $v_2$ at $p_\mathrm{T}\sim 15$~GeV as compared to the results presented in this work with only collisional energy loss. The final goal of Catania is to reach a new step forward that includes multiple gluon emission processes in a model where the radiation mechanism can be consistently coupled to non-perturbative effects, as discussed in the QPM model and proven to play an important role for HQ thermalization. 

The TAMU group is planning to extend the non-perturbative $T$-matrix approach to include gluon radiation. In addition, the TAMU group plans to implement a viscous (instead of ideal) hydrodynamic evolution including fluctuating initial conditions. The PHSD approach plans to include the radiative energy loss of heavy-quarks as well which allows us to extend the results to larger $p_\mathrm{T}$. 

For the elastic energy loss of heavy quarks, the present CCNU-LBNL and Nantes approaches are based on perturbative collisions of heavy quarks with massless partons that constitute the QGP. Both these two models plan to introduce the finite thermal mass of light partons as implemented in the current Catania, TAMU and PHSD models. This is important to allowing the thermal parton distribution to respect the lattice equation-of-state that is applied in the hydrodynamical evolution. In addition, the CCNU-LBNL model plans to extend its treatment of the elastic scattering process beyond the $2\rightarrow 2$ perturbative description by introducing the heavy-quark-potential interaction as established in the TAMU model. 

In the current Nantes model, the heavy quarks are produced at the initial interaction points of the incoming baryons (which are eventually part of a heavy ion). Their transverse momentum distribution is given by FONLL while some shadowing is modeled through standard nPDF. The Nantes group is currently working on EPOS-HQ that is based on the EPOS3 model. In this approach the heavy quarks are commonly produced with the light partons, sharing with them the energy, so the initial distribution has to come out in a more consistent way within the framework and not as an external input. Furthermore, the medium modification of the parton distribution function, like shadowing, are automatically taken into account. The Nantes group also plans to improve the description for heavy quarks with a large transverse momentum by better dealing with the gluon formation-time in radiative energy loss. 

Last but not least, the Duke group will combine separate treatments of soft and hard medium-probe interactions into a unified approach, allowing an interpolating description between soft diffusion and hard collisions. Meanwhile, it plans to improve the current LPM implementation to better agree with theoretical calculations. These developments are expected to aid the future quantification of heavy quark transport properties with a better constrained uncertainty.

\section{Summary and outlook}
\label{sec:summary}

We have carried out a systematic and comparative study of six different transport models for HF meson production in heavy-ion collisions. While all models have passed the basic consistency check in calculating the transport coefficients with pQCD Born diagrams for elastic scattering, the extracted HQ transport coefficients with the parameters chosen to reproduce the experimental heavy-ion data at LHC differ by a up to a factor of 5(3) at high (small) momenta between different models. 

To study whether and how these differences are consequences of different treatments of physical processes in these approaches, for example, the hadronization or the expansion of the QGP,  we have eliminated the latter two by designing a simple static QGP brick medium with a fixed temperature and length that mimics HQ propagation in central Pb+Pb collisions. By adjusting the parameters in each model (tune 2) to give a fixed value of the HQ suppression factor for a given initial transverse momentum of $p_\mathrm{T}=15$ GeV/$c$, the differences of the HQ drag coefficient from different models are reduced to a factor of 2. This implies that different bulk medium evolutions and HQ hadronization have a substantial influence on HF meson suppression in heavy-ion collisions. In the tune 2 calculations,  we observe that the numerical values of the transport coefficients from different models fall into three different groups: models based on elastic HQ scattering off thermal quasi-particles with large masses, especially near the phase transition, models with elastic scatterings off partons with moderate quasi-particle masses, and models that include both elastic and inelastic collisions. This indicates that the remaining differences in the numerical values of the  transport coefficients among the three groups of models can be attributed to the treatment of elastic and inelastic HQ interactions in the medium, as well as to the masses of the quark and gluon quasi-particles in the QGP. In addition, different treatments of HQ formation times and the transport schemes (Langevin vs. Boltzmann as summarized in Tab.~\ref{tab:tuning}) by different models introduce additional sources of the remaining discrepancy~\cite{Das:2013kea}.

Assuming that the initial momentum distribution of heavy quarks can be calculated with pQCD, the physics of heavy quarks in an expanding QGP is determined mainly by three processes: The expansion of the QGP fluid, the interaction of heavy quarks with the QGP constituents, and their hadronization. Present experimental results do not allow us to decisively separate these components. Further progress is possible in both the theoretical and phenomenological directions.  On the theoretical side, the present study identified large uncertainties arising from the modeling of the bulk evolution. Therefore, systematic and comparative studies of the models for bulk evolution should be carried out with constraints from the experimental data on bulk hadron spectra. Substantial progress has been made in this direction in recent years which should be incorporated into the study of HQ transport phenomena.  Thus, as the next step of our collaborative effort, we will implement different models of heavy quark medium interaction within a common realistic hydrodynamic medium that has been well constrained by the soft hadron observables, which is crucial to minimizing the systematic uncertainty of the extracted heavy quark transport coefficient in a realistic QGP medium.

Additionally, inelastic interactions such as induced gluon radiation have been studied in detail in the past two decades and are found to be responsible for jet quenching observed in experiments at RHIC and the LHC. They should be incorporated into all theoretical models for the transport of high-momentum heavy quarks through the QGP. For elastic interactions, ample constraints are available from lattice-QCD data which are particularly relevant at low and intermediate HQ momenta, including hadronization processes. 

On the phenomenological side, new experimental measurements such as angular correlations between $D$ mesons and light hadrons or $D D(\bar D)$ correlations can provide further guidance for theoretical models, in particular on the mass scale of the thermal quasi-particles which affect the angular distribution of HQ-parton scattering. Recent experimental data in high multiplicity p+p and p+A collisions indicate the formation of a QGP and collective phenomena in small systems. The study of the modification of HF meson spectra in these small systems can also help to elucidate the nature of HQ interactions in medium and the approach to thermalization. Therefore, within a theory collaboration, we will also conduct a systematic assessment of the underlying physics in each model with constraints provided by additional experimental observables such as heavy hadron anisotropy, heavy-light (heavy) hadron correlations, heavy-quark jet shape and fragmentation functions.  All of the above are necessary steps to take toward reducing the theoretical and phenomenological uncertainties in the extraction of the heavy-quark transport coefficients in the QGP.

\section*{Acknowledgments}

This work was initiated and initially supported by the JET Collaboration. We thank M. Gyulassy and G.-Y. Qin for helpful discussions. This work was supported by the Director, Office of Energy Research, Office of High Energy and Nuclear Physics, Division of Nuclear Physics, of the U.S. Department of Energy (DOE) under Grants No. DE-AC02-05CH11231 (JET) (X.-N.W.), No. DE-SC0013460 (S.C.), and No. DE-FG02-05ER41367 (S.B., W.K., Y.X.); the U.S. National Science Foundation (NSF) under Grants No. ACI-1550228 (JETSCAPE) (X.-N.W.), No. ACI-1550300 (JETSCAPE) (S.C.), No. ACI-1550225 (JETSCAPE) (S.B., W.K.), No. PHY-1306359 (M.H., R.R.) and No. PHY-1614484 (M.H., S.L., R.R.); the National Science Foundation of China (NSFC) under Grants No. 11221504 (X.-N.W.), and No. 11675079 (M.H.); the Major State Basic Research Development Program in China under Grant No. 2014CB845404 (X.-N.W.); the R\'{e}gion Pays de la Loire (France) under Contract No. 2015-08473 (J.A., P.B.G, M.N.); the German Academic Exchange Service (DAAD) (T.S., E.B.); the Deutsche Forschungsgemeinschaft (DFG) under Grant No. CRC-TR 211 (E.B.); and the European Research Council (ERC) StG under Grant No. 259684 (G.C., S.K.D, V.G, S.P., F.S.).

\bibliographystyle{h-physrev5}
\bibliography{SCrefs}

\end{document}